\documentclass[12pt]{article}
\usepackage{a4wide}
\usepackage{amssymb}
\usepackage{graphicx}
\usepackage{subfigure}
\usepackage{epsfig}
\usepackage[tbtags]{amsmath}
\usepackage{exscale}
\usepackage{appendix}

\begin{document}

\newcommand{\kp}{\ensuremath{K_\varphi}}
\newcommand{\kx}{\ensuremath{K_x}}
\newcommand{\ex}{\ensuremath{E^x}}
\newcommand{\ep}{\ensuremath{E^\varphi}}
\newcommand{\gp}{\ensuremath{\Gamma_\varphi}}
\newcommand{\be}{\begin{equation}}
\newcommand{\ee}{\end{equation}}
\newcommand{\bea}{\begin{eqnarray}}
\newcommand{\eea}{\end{eqnarray}}
\newcommand{\vd}[2]{\frac{\delta #1}{\delta #2}}   %define variational derivative   
\newcommand{\Ef}{E^\varphi}
\newcommand{\Kf}{K_\varphi}
\newcommand{\lP}{\ell_{\mathrm P}}

\newcommand{\md}{{\mathrm d}}

\begin{center}
{\Large Spherically symmetric Einstein-Maxwell theory and loop quantum gravity corrections} \\
\vspace{1.5em}
Rakesh Tibrewala \footnote{e-mail address: {\tt rtibs@iisertvm.ac.in}}
\\
\vspace{1em}
School of Physics, Indian Institute of Science Education and Research,
CET Campus, Trivandrum 695016, India
\end{center}

\begin{abstract}
Effects of inverse triad corrections and (point) holonomy corrections, occuring in loop quantum gravity, are considered on the properties of Reissner-Nordstr\"om black holes. The version of inverse triad corrections with unmodified constraint algebra reveals the possibility of occurrence of three horizons (over a finite range of mass) and also shows a mass threshold beyond which the inner horizon disappears. For the version with modified constraint algebra, coordinate transformations are no longer a good symmetry. The covariance property of spacetime is regained by using a \emph{quantum} notion of mapping from phase space to spacetime. The resulting quantum effects in both versions of these corrections can be associated with renormalization of either mass, charge or wave function. In neither of the versions, Newton's constant is renormalized. (Point) Holonomy corrections are shown to preclude the undeformed version of constraint algebra as also a static solution, though time-independent solutions exist. A possible reason for difficulty in constructing a covariant metric for these corrections is highlighted. Furthermore, the deformed algebra with holonomy corrections is shown to imply signature change.     
\end{abstract}

\section{Introduction}
Symmetries play a fundamental role in physics. Certain symmetries, like the spacetime symmetries, lead to familiar conservation laws for energy and momentum. There are also gauge symmetries familiar in particle physics (which actually reflect the redundancy of the classical formulation of a physical theory) and are of prime importance, especially in the context of quantum theory. In a diffeomorphism covariant theory like general relativity, these two notions of symmetry become closely related. This is particularly clear when contrasting the usual action formulation of general relativity with the canonical Arnowitt-Deser-Misner (ADM) formulation \cite{adm}. 

While the spacetime covariance property of the theory is obvious in the standard tensorial formulation, in the ADM formulation spacetime symmetry gives way to gauge symmetry in the phase space and is encoded in the form of constraint algebra of the theory. Gauge transformations generated by the constraints then correspond to spacetime diffeomorphisms. It then becomes important to ask: What happens to the standard covariance property of the theory in the quantum domain? Does the symmetry survives in the quantum theory or it becomes anomalous or is it just deformed? In loop quantum gravity, where the ADM variables are replaced by Ashtekar variables \cite{ashtekarnewvariables}, the issue has recently seen renewed interest \cite{martinpaily1, tensorialspacetimes, u1cubemodel}. 

In loop quantum cosmology, where the program of loop quantum gravity has found its most successful application \cite{martinLQC, martinLQCreview, ashtekarparam}, the above issue trivializes since the diffeomorphism constraint is identically equal to zero. Addition of perturbative inhomogeneities makes a full quantum analysis difficult but at an effective level, where inverse triad corrections or holonomy corrections are included, it shows that the constraint algebra is deformed \cite{martingolamshanki}. 

Like cosmological models, spherically symmetric models have always been a very useful test bed for theories of quantum gravity and these have been studied within loop quantum gravity for different applications (see for instance \cite{gambinipullin, viqar, modesto}). These models are interesting for several reasons. Black hole geometries in classical general relativity are mostly spherically symmetric and thus provide an opportunity to study the possible effects of quantum gravity on issues related to singularity and horizon properties. %For instance in \cite{massthreshold, modifiedhorizon} it has been found that there is a mass threshold below which the black hole horizon disappears. 
Modifications to horizon properties can, for instance, have consequences for Hawking radiation, particularly, during the end stage of black hole evaporation. These models are quite versatile and apart from allowing an exploration of black hole spacetimes which mostly have time-independent geometry, they also allow time-dependent gravitational collapse scenarios \cite{viqar critical, LTB1, LTB2, ziprick effective action, viqar mass gap, viqar modified GR}.
% which have also been looked at from the perspective of loop quantum gravity in \cite{LTB1, LTB2}. 

Apart from these practical considerations there is another very important reason for the usefulness of these models. As opposed to cosmological models, the constraint algebra is non-trivial here and as pointed out in the beginning, it is the constraint algebra which encodes the symmetry properties of the theory. These models therefore provide an opportunity to study the effects of quantum gravity on the symmetries of the classical theory without entailing all the difficulties that a full theory without any symmetry reduction would have. Moreover, the inhomogeneity is non-perturbative implying (in certain sense) a closer association with the full theory. Like cosmological models incorporating perturbative inhomogeneities, in several of these models, loop quantum gravity inspired corrections lead to a deformed constraint algebra \cite{LTB2, PoissonSigma, modifiedhorizon}.

Implications of the classical hypersurface deformation algebra from the perspective of the canonical theory have been analyzed in \cite{hojman} and its implication for the Lagrangian of the theory in \cite{geometrodynamicslagrangian}. Recently, in \cite{martinpaily1, martinpaily2}, the approach of \cite{geometrodynamicslagrangian} was adopted for the case of modified hypersurface deformation algebra and it highlighted several interesting features including generic singularity resolution as a direct consequence of the deformation of the constraint algebra. 

While the utility of such general analyses is obvious, it is also useful to consider specific models since they allow explicit solutions to be obtained which can be compared and contrasted with their classical counterparts and can thus provide useful hints for more generic constructions. Such an analysis was performed for spherically symmetric vacuum spacetimes and those incorporating perturbative scalar matter in \cite{modifiedhorizon} (see \cite{icgcproceedings} for a summary of some of the results). It was found that with deformed algebra the familiar spacetime concepts (spacetime metric, for instance) need to be reconsidered as also the constructs like the black hole horizon, which are based on such notions.
    
In this paper, we extend the work in \cite{modifiedhorizon} and consider minimally coupled Einstein-Maxwell theory with spherically symmetric background. Reduction to spherical symmetry in terms of Ashtekar variables has been discussed in \cite{thiemSphSymm1, thiemSphSymm2, martinkastrup} and including Maxwell field in \cite{thiemann eins-max} (also see \cite{dategolam}). The aim of the present work is to include loop quantum gravity corrections at an effective level and to work out the consequences for the Reissner-Nordstr\"om black hole. The main theme would be to maintain the first class nature of the constraint algebra (though it can be deformed). 

In \cite{modifiedhorizon}, only inverse triad corrections arising from loop quantum gravity were considered. Here, we also consider the effects of (point) holonomy correction, and unlike cosmology, the matter field can be non-perturbative. Since the details of the theory for these more complicated models are still unclear, it is useful to explore various models at an effective level incorporating different matter fields. Holonomy corrections have so far not been analyzed in much detail for spherically reduced models though they have been discussed in some detail in \cite{juanthesis}. These corrections can lead to some very dramatic effects leading to the signature change of the metric as has been pointed out recently in \cite{grain signature, martinpaily1}. 

At the beginning itself, we would like to clarify that whereas the form of the inverse triad corrections can be derived from the underlying spherically symmetric loop quantum gravity (in that the operator realization of inverse triad factors can be obtained by following the formal procedures of the theory and their eigenvalues evaluated on the spherically symmetric spin network states, see \cite{LTB1} for a derivation), the form of the point holonomy correction has not been derived but is only motivated from the form of similar corrections in loop quantum cosmology. 

With quantum-corrected constraints and constraint algebra, it is not guaranteed that a consistent set of equations will result. We will show that at the effective level considered here, it is possible to obtain consistent equations for both inverse triad and holonomy corrections. Since we want to include the effects of loop quantization which requires a Hamiltonian framework, we start in section 2 with the Hamiltonian formulation of the classical theory in terms of Ashtekar variables. After performing the constraint analysis at the classical level, we will move on to consider the effects of inverse triad corrections in section 3. As in earlier works, we will consider two cases - inverse triad corrections which keep the constraint algebra unmodified and those which modify the algebra. By a modified or a deformed constraint algebra we mean that the structure function on the r.h.s. of the Poisson bracket(s) involving the constraints is different from what one would get in classical general relativity. For the class of corrections considered here, it will turn out that only the bracket involving two Hamiltonians is modified. 

We then solve the constraints and the equations of motion for the phase space variables. For the case where the constraint algebra is unmodified, we show that the usual procedure of obtaining a spacetime metric from phase space variables leads to a physically meaningful object. By this we mean that one can perform coordinate transformation on this solution to obtain the metric in different coordinates, and this metric, obtained after coordinate transformation, still solves the constraints and the equations of motion. 

In contrast, for the case where the constraint algebra is modified, even though we can write a metric following the usual procedure, it turns out that such an object is not physically meaningful. We will explicitly show that in this case if one performs a coordinate transformation on this \emph{supposed} spacetime metric to obtain the metric in new coordinates, the transformed metric does not solve the system consisting of the constraint equations and the equations of motion. That is, for modified algebra, the familiar spacetime covariance property, which, under coordinate transformation, would map solutions of constraints to other solutions, does not hold. To recover such a notion, we provide a \emph{quantum} version of mapping from phase space to spacetime metric and show that with this new mapping the spacetime covariance property is recovered such that under coordinate transformation the solutions of constraints and of equations of motion are mapped to other solutions. 

In section 4, we will consider the inclusion of corrections due to point holonomies corresponding to spherically symmetric directions. Here also we will consider two cases: (1) phase-space-independent holonomy corrections and (2) phase-space-dependent holonomy corrections. By phase-space-dependent or phase-space-independent holonomy corrections, we mean whether the scale associated with these corrections (denoted $\delta$ in the text) is a function of the phase space variables (specifically the triad variables) or not. 

The effect of the inclusion of holonomy correction turns out to be non-trivial and it seems that a static solution or a solution with unmodified constraint algebra is inconsistent with the inclusion of holonomy corrections in either form. Furthermore, unlike the case of inverse triad corrections, here it turns out to be much more difficult to regain the usual notion of covariance for the solution of constraints and equations of motion. We point out a possible reason for the difficulty and finally conclude in section 5. We will work with metric signature $(-+++)$ and unless otherwise specified the Greek indices $(\mu, \nu...)$ will denote spacetime indices taking values $(0,1,2,3)$. 

\section{Classical theory}

The system we are interested in is the minimally coupled Einstein-Maxwell theory with spherical symmetry and is described by the action
\be \label{action}
S=\int \md^{4}x(\mathcal{L}_{G}+\mathcal{L}_{EM})=\int \md^{4}x\sqrt{-g}\left(\frac{1}{16\pi G}R-\frac{1}{4}F_{\mu\nu}F^{\mu\nu}\right),
\ee
where $R$ is the Ricci scalar, $F_{\mu\nu}$ is the electromagnetic field strength, $g$ is the determinant of the metric tensor $g_{\mu\nu}$ and $G$ is Newton's constant. To investigate the effects of loop quantum gravity, we need the canonical formulation of the above theory which for the minimally coupled case can be performed independently for the gravitational and the electromagnetic sector. We start with the gravitational sector where, since we are interested in loop quantization, we use densitized triads
instead of the spatial metric components. For spherical symmetry, using the $\mathfrak{su}(2)$ basis $\tau_i$ this is (see \cite{SymmRed,SphSymmstates,SphSymmHam} for details): 
\be
E=E^x(x)\tau_3\sin\theta\frac{\partial}{\partial x}+
(E^1(x)\tau_1+E^2(x)\tau_2)\sin\theta\frac{\partial}{\partial\theta} 
+(E^1(x)\tau_2-E^2(x)\tau_1)\frac{\partial}{\partial\phi}. \notag
\ee
Momenta conjugate to the triad variables are given in terms of the Ashtekar-Barbero connection $\mathcal{A}^{i}_{a}=\Gamma^{i}_{a}+\gamma
K^{i}_{a}$, where $\Gamma^{i}_{a}$ and $K^{i}_{a}$ are the components of spin connection and extrinsic curvature, respectively, and $\gamma$
is the Barbero-Immirzi parameter \cite{barbero,immirzi}. For spherical symmetry 
\be
\mathcal{A}=\mathcal{A}_x(x)\tau_3\md x+(\mathcal{A}_1(x)\tau_1+\mathcal{A}_2(x)\tau_2)\md\theta
+(\mathcal{A}_1(x)\tau_2-\mathcal{A}_2(x)\tau_1)\sin\theta\md\phi+ \tau_3\cos\theta\md\phi.
\ee
If one introduces the U(1)-gauge invariant quantities
$(E^{\varphi})^2=(E^1)^2+(E^2)^2$ and $\mathcal{A}_{\varphi}^2=\mathcal{A}_1^2+\mathcal{A}_2^2$, then the symplectic structure is
\begin{align}
\{\mathcal{A}_x(x),E^x(y)\}&=\{\gamma K_\varphi(x),2E^\varphi(y)\} \notag\\
&=\{\eta(x),P^\eta(y)\}=2G\gamma \delta(x,y).  \notag
\end{align}
%or more explicitly the Poisson bracket of functions $f$ and $g$ is
%\begin{align}
%\{f,g\}=&2G\int \md x\bigg(\gamma\vd{f}{\mathcal{A}_x}\vd{g}{E^x}+\frac{1}{2}\vd{f}{\Kf}\vd{g}{\Ef}+\gamma\vd{f}{\eta}\vd{g}{P^\eta}  \notag\\
%             &-\gamma\vd{f}{E^x}\vd{g}{\mathcal{A}_x}-\frac{1}{2}\vd{f}{\Ef}\vd{g}{\Kf}-\gamma\vd{f}{P^\eta}\vd{g}{\eta}\bigg)\,. \notag
%\end{align}
Here the field $\eta(x)$ is a U(1)-gauge angle and has the conjugate momentum
\begin{align}
 P^{\eta}(x)&=2\mathcal{A}_{\varphi}E^{\varphi}\sin\alpha \notag\\
 &=4{\rm tr}\left((E^1\tau_1+E^2\tau_2)(\mathcal{A}_2\tau_1-\mathcal{A}_1\tau_2)\right) \notag
\end{align}
(with $\alpha$ defined as the angle between the internal directions of $\mathcal{A}$- and $E$-components). With this the $x$-component of the spin connection turns out to be $\Gamma_x=-\eta'$ implying that the corresponding Ashtekar connection is $\mathcal{A}_x=-\eta'+\gamma K_x$ (where $K_{x}$ is the
extrinsic curvature component).

In terms of these variables we have the Gauss constraint
\begin{equation}
G_{\rm grav}[\lambda]=\frac{1}{2G\gamma}\int \md x\,
\lambda((E^x)'+P^\eta),  \label{Gauss}
\end{equation}
the vector constraint
\begin{align}
D_{\rm grav}[N^x]=\frac{1}{2G}\int &\md x\,N^x\bigg(2E^\varphi K_\varphi' 
-\frac{1}{\gamma}\mathcal{A}_x(E^x)'+\frac{1}{\gamma}\eta'P^\eta\bigg) \notag\\
=\frac{1}{2G}\int &\md x\,N^x\bigg(2E^\varphi K_\varphi'-K_x(E^x)'
+\frac{1}{\gamma}\eta'((E^x)'+P^\eta)\bigg)      \label{grav vector}
\end{align}
and the gravitational part of the Hamiltonian constraint 
\be
H_{\rm grav}[N]=-\frac{1}{2G}\int \md x\,N|E^x|^{-\frac{1}{2}}(K_\varphi^2E^\varphi+2K_\varphi K_xE^x
+ (1-\Gamma_\varphi^2)E^\varphi+2\Gamma_\varphi'E^x)  \label{Hamclass}
\ee
with $\Gamma_\varphi=-(E^x)'/2E^\varphi$ being the gauge-invariant angular
component of the spin connection. The spherically symmmetric metric in terms of these variables is given by
\be \label{sphsymm adm metric}
\md s^{2}=-N^{2}\md t^{2}+\frac{(\ep)^{2}}{\ex}(\md x+N^{x}\md t)^{2}+\ex \md \Omega^{2}.
\ee
Here $N$ and $N^{x}$ are the lapse function and the shift vector, respectively, and turn out to be the Lagrange multipliers of the theory, and $\ep$ and $\ex$ are the dynamical variables. The radial direction is coordinatized by $x$, and $\md \Omega^{2}=\md\theta^{2}+\sin^{2}\theta \md\phi^{2}$ is the angular part of the metric.  In terms of these variables, the determinant of the four metric is given by $\sqrt{-g}=N\ep\sqrt{\ex}\sin\theta$.

We now move on to the Maxwell field for which the Lagrangian density is 
\be \label{maxwell lagrangian}
\mathcal{L}_{EM}=-\frac{1}{4}\sqrt{-g}F_{\mu\nu}F^{\mu\nu},
\ee
where the electromagnetic field tensor $F_{\mu\nu}=\partial_{\mu}A_{\nu}-\partial_{\nu}A_{\mu}$ in terms of the electromagnetic four potential $A_{\mu}$. Momentum conjugate to $A_{\mu}$ is 
\be \label{em momenta}
\pi^{\mu}=\frac{\delta\mathcal{L}_{EM}}{\delta\dot{A}_{\mu}}=\sqrt{-g}F^{\mu 0}.
\ee 
Since $F^{\mu\nu}$ is antisymmetric, we have the well known result $\pi^{0}=0$ or in the smeared form
\be \label{primary constraint}
C_{1}[\nu]=\int \md x\,\nu(x)\pi^{0}(x)\approx0
\ee
which is the primary constraint of the theory (and where, in writing the above expression, we have already imposed spherical symmetry).  
%We are interested in the Hamiltonian of the theory and for this following the Arnowitt-Deser-Misner (ADM) approach, we perform a $3+1$ splitting of the spacetime. We specialize to spherically symmetric backgrounds and keeping in mind the application to a loop quantization choose to work with the su(2) or Ashtekar variables (a discussion of the spherically symmetric Ashtekar variables will be presented below)  

Before working out the Hamiltonian, we note that for spherical symmetry there will not be any $\theta$ and $\phi$ components of the electric and magnetic fields. We will further assume that there are no magnetic fields. This can be achieved by having the Lagrangian depend only on $A_{x}$ and $A_{t}$ components of the four potential, with both these fields depending only on $(t,x)$. In such a situation, only the radial component of $\pi^{\mu}$ will be non-zero:
\be \label{radial momenta}
\pi^{x}(t,x)=N\ep\sqrt{\ex}F^{x0}\sin\theta=p^{x}\sin\theta
\ee
where, in the above equation, we have introduced the notation $p^{x}=N\ep\sqrt{\ex}F^{x0}$. The fields $A_{x}$ and $\pi^{x}$ obey the Poisson bracket relation $\{A_{x}(\vec{y}),\pi^{x}(\vec{z})\}=\delta^{3}(\vec{y},\vec{z})$ or after integrating over the angular directions we have the relation
\be \label{em poisson bracket}
\{A_{x}(y),p^{x}(z)\}=\frac{1}{4\pi}\delta(y,z)
\ee
where $\vec{y}$ denotes all three spatial coordinates and $y$ denotes only the radial coordinate. With $\pi^{0}(x)=0$, the Hamiltonian 
\[
\mathcal{H}_{EM}=\int \md^{3}y\left[\pi^{x}(y)\dot{A}_{x}(y)-\mathcal{L}_{EM}\right],
\]
after integrating over the angular coordinates is given by
\be 
\mathcal{H}_{EM}=4\pi\int \md x\left[\frac{N\ep(p^{x})^{2}}{2(\ex)^{3/2}}+p^{x}\partial_{x}A_{0}\right].
\ee
We can perform an integration by parts on the last term and write the Hamiltonian as
\be \label{matter hamiltonian}
\mathcal{H}_{EM}=4\pi\int \md x\left[\frac{N\ep(p^{x})^{2}}{2(\ex)^{3/2}}-A_{0}\partial_{x}p^{x}\right].
\ee

We already have a primary constraint \eqref{primary constraint} in the theory, and following Dirac \cite{dirac}, we need to work out the consistency condition(s) that this implies. Since this constraint does not involve any other canonical variable apart from $\pi^{0}\equiv p^{0}\sin\theta$, we can neglect the gravitational part of the Hamiltonian and calculate the Poisson bracket  $\{p^{0},\mathcal{H}_{EM}\}$ which, using $\{A_{0}(y),p^{0}(z)\}=\delta(y,z)/4\pi$, is
\be
\{p^{0}(x),\mathcal{H}_{EM}\}=\{p^{0}(x),4\pi\int \md y\left[\frac{N\ep(p^{x})^{2}}{2(\ex)^{3/2}}-A_{0}\partial_{y}p^{x}\right]\}=p^{x'}(x),
\ee
where prime ($'$) denotes derivative with respect to $x$. For consistency, this should equal zero, implying that we have a secondary constraint in the theory $p^{x'}\approx0$, which we write in the smeared form as
\be \label{secondary constraint}
C_{2}[\beta]=\int \md x\,\beta(x)p^{x'}(x)\approx0.
\ee
This would be recognized as the Gauss constraint of the electromagnetic theory. We also need to evaluate the Poisson bracket of this secondary constraint with the Hamiltonian to see if it gives any further conditions. It is easily checked that this Poisson bracket does not give any new condition and turns out to be identically zero (since $\mathcal{H}_{EM}$ does not depend on $A_{x}$). We also note that the matter Hamiltonian $\mathcal{H}_{EM}$ itself involves the constraint $C_{2}[4\pi A_{0}]$ which we separate out and write the electromagnetic Hamiltonian as 
\be
\mathcal{H}_{EM}=H_{EM}[N]-C_{2}[4\pi A_{0}], \nonumber
\ee
where 
\be \label{em hamiltonian}
H_{EM}[N]=4\pi\int \md x\left(\frac{N\ep(p^{x})^{2}}{2(\ex)^{3/2}}\right).
\ee 
%We now have five constraints in the theory \eqref{Gauss}, \eqref{grav vector}, \eqref{primary constraint}, \eqref{secondary constraint} and the Hamiltonian constraint
%\bea \label{ham constraint}
%H[N]&=&-\frac{1}{2G}\int \md x N|E^x|^{-\frac{1}{2}}(K_\varphi^2E^\varphi+2K_\varphi K_xE^x 
%+ (1-\Gamma_\varphi^2)E^\varphi+2\Gamma_\varphi'E^x) \nonumber \\
%&&+4\pi\int \md x\left(\frac{N\ep(p^{x})^{2}}{2(\ex)^{3/2}}\right).
%\eea
%(For later purpose (dealing with inverse triad corrections) we note that in the matter part of the Hamiltonian $\ex$ comes with a power of $-3/2$ while in the gravitational part it comes with a power of $-1/2$.)

Note that due to the antisymmetric nature of the Maxwell field and due to the imposition of spherical symmetry with $A_{\theta}=0=A_{\phi}$, there is no matter contribution to the vector constraint. Working directly from the action we only obtain the vector constraint. It is, however, obvious that matter fields should transform under radial diffeomorphisms. To get the matter contribution to the diffeomorphism constraint, we turn to \cite{geometrodynamicslagrangian} where it has been shown that we can obtain the diffeomorphism constraint directly from the requirement that the transformation generated by it should be equal to the corresponding Lie derivative:
\be
\{F,D\}\delta N^{x}=\mathcal{L}_{\overrightarrow{\delta N}^{x}}F.
\ee
Specifically, for the spherically symmetric Maxwell field, where $A_{x}$ is a one-form and $p^{x}$ is a scalar, this implies (using \eqref{em poisson bracket}):
\bea
\frac{1}{4\pi}\frac{\delta D_{EM}(y)}{\delta p^{x}(x)}\delta N^{x}(y) &=& A_{x,x}\delta N^{x}(x)+A_{x}(x)\delta N^{x}(x)_{,x}, \nonumber \\
-\frac{1}{4\pi}\frac{\delta D_{EM}(y)}{\delta A_{x}(x)}\delta N^{x}(y) &=& p^{x}(x)_{,x}\delta N^{x}(x).
\eea
It is easy to check that the above equations are satisfied for $D_{EM}(x)=-4\pi\int\md x A_{x}(x)p^{x'}(x)$ which is the matter contribution to the diffeomorphism constraint. We therefore add this contribution to the vector constraint \eqref{grav vector} to get the diffeomorphism constraint (which we continue to denote as $D$)
\be \label{diffeomorphism}
D[N^x]=\frac{1}{2G}\int \md x\,N^x\bigg(2E^\varphi K_\varphi'-K_x(E^x)'
+\frac{1}{\gamma}\eta'((E^x)'+P^\eta)\bigg)-4\pi\int \md x\,N^{x}A_{x}(x)p^{x'}(x)
\ee

We now have five constraints in the theory \eqref{Gauss}, \eqref{primary constraint}, \eqref{secondary constraint}, \eqref{diffeomorphism} and the Hamiltonian constraint
\bea \label{ham constraint}
H[N]&=&-\frac{1}{2G}\int \md x N|E^x|^{-\frac{1}{2}}(K_\varphi^2E^\varphi+2K_\varphi K_xE^x 
+ (1-\Gamma_\varphi^2)E^\varphi+2\Gamma_\varphi'E^x) \nonumber \\
&&+4\pi\int \md x\left(\frac{N\ep(p^{x})^{2}}{2(\ex)^{3/2}}\right).
\eea
(For later purpose dealing with inverse triad corrections, we note that in the matter part of the Hamiltonian $\ex$ comes with a power of $-3/2$, while in the gravitational part it comes with a power of $-1/2$.)

Next, we verify that the set of five constraints forms a first class system. Before doing that it is useful to note that $C_{1}[\nu]$ and $C_{2}[\beta]$ depend only on $p^{0}$ and $p^{x}$, respectively, and, in particular, are independent of gravitational fields. Similarly, $G[\lambda]$ is independent of Maxwell fields and the matter dependence of $D[N^{x}]$ in only through the pair $(A_{x},p^{x})$. This then immediately implies that $C_{1}[\nu]$ commutes with all the constraints.
%\bea \label{algebra for c1}
%\{C_{1}[\nu],C_{1}[\overline{\nu}]\}&=&0 \nonumber \\
%\{C_{1}[\nu],C_{2}[\beta]\}&=&0 \nonumber \\
%\{C_{1}[\nu],G[\lambda]\}&=&0 \nonumber \\
%\{C_{1}[\nu],D[N^{x}]\}&=&0 \nonumber \\
%\{C_{1}[\nu],H[N]\}&=&0
%\eea
Similarly, we find that $C_{2}[\beta]$ Poisson commutes with all the constraints except the diffeomorphism constraint with which it gives
\be
\{D[N^{x}],C_{2}[\beta]\}=C_{2}[N^{x}\beta'].
\ee
Earlier, in working out the consistency condition for the secondary constraint in \eqref{secondary constraint}, we considered its Poisson bracket only with $\mathcal{H}_{\rm EM}$ since at that stage we had not included the matter contribution towards the diffeomorphism constraint. The above Poisson bracket shows that had we also included the matter contribution to diffeomorphism constraint, we would not have got any extra consistency condition.

The gravitational Gauss constraint $G[\lambda]$ Poisson commutes with itself ($G[\overline{\lambda}]$) and with the Hamiltonian constraint. Its Poisson bracket with the diffeomorphism constraint is
\be
\{G[\lambda],D[N^{x}]\}=-G[\lambda'N^{x}].
\ee
For the diffeomorphism constraint, we have $\{D[N^{x}],D[M^{x}]\}=D[N^{x}M^{x'}-N^{x'}M^{x}]$. To evaluate $\{D[N^{x}],H[N]\}$, we write $H[N]=H_{\rm grav}[N]+H_{EM}[N]$ and find that
\bea
\{D[N^{x}],H_{\rm grav}[N]\} &=& H_{\rm grav}[N'N^{x}], \nonumber \\
\{D[N^{x}],H_{EM}[N]\} &=& H_{EM}[N'N^{x}], \nonumber 
\eea
which implies
\be
\{D[N^{x}],H[N]\}=H[N^{x}N']. 
\ee
Finally, we have to check the Poisson bracket of the Hamiltonian constraint with itself -- $\{H[N],H[M]\}$. To evaluate this, as before, we write $H[N]=H_{\rm grav}[N]+H_{EM}[N]$ and note that $\{H_{EM}[N],H_{EM}[M]\}=0$, $\{H_{\rm grav}[N],H_{EM}[M]\}=-\{H_{EM}[N],H_{\rm grav}[M]\}$ and therefore the only non-trivial contribution comes from the $\{H_{\rm grav}[N],H_{\rm grav}[M]\}$ bracket and we find
\be \label{classical hh bracket}
\{H[N],H[M]\}=D[|\ex|(\ep)^{-2}(NM'-N'M)]-G[|\ex|(\ep)^{-2}(NM'-N'M)\eta'].
\ee
Thus the constraints form a first class system. At this stage, we can simplify the analysis somewhat by solving the gravitational Gauss constraint $G[\lambda]$ and the primary constraint $C_{1}[\nu]$. This removes the pairs $(\eta,P^{\eta})$ and $(A_{0},\pi^{0})$ and we are left with the canonical pairs
\be
\{K_x(x),E^x(y)\}=\{K_\varphi(x),2E^\varphi(y)\}=2G \delta(x,y) \quad {\rm and} \quad \{A_{x}(x),p^{x}(y)\}=\frac{1}{4\pi}\delta(x,y).
\ee
Among the remaining three constraints, while $C_{2}[\beta]$ and $H[N]$ retain their form (remembering that now $K_{x}$ is one of the canonical variables) the diffeomorphism constraint reduces to
\be \label{diffeo}
D[N^{x}]=\frac{1}{2G}\int \md x\,N^{x}(2\Kf'\ep-\kx E^{x'})-4\pi\int \md x\,N^{x}A_{x}(x)p^{x'}(x).
\ee

\subsection*{Equations of motion}
The total Hamiltonian $H_{T}=H[N]+D[N^{x}]+C_{2}[\beta]$ and using it we can work out the equations of motion for the dynamical variables of the theory using Hamilton's equations $\dot{a}=\{a,H_{T}\}$. 
\bea \label{classical eom}
\label{ex dot}
\dot{E}^{x} &=& N^{x}E^{x'}+2N\kp\sqrt{\ex} \\
\label{ephi dot}
\dot{E}^{\varphi} &=& (N^{x}\ep)'+\frac{N\kp\ep}{\sqrt{\ex}}+N\kx\sqrt{\ex} \\
\label{kphi dot}
\dot{K}_{\varphi} &=& N^{x}\kp'-\frac{N\kp^{2}}{2\sqrt{\ex}}-\frac{N}{2\sqrt{\ex}}+\frac{N(E^{x'})^{2}}{8(\ep)^{2}\sqrt{\ex}}+\frac{N'E^{x'}\sqrt{\ex}}{2(\ep)^{2}}+2\pi G\frac{N(p^{x})^{2}}{(\ex)^{3/2}} \\
\label{kx dot}
\dot{K}_{x} &=& (N^{x}\kx)'+\frac{N\kp^{2}\ep}{2(\ex)^{3/2}}-\frac{N\kp\kx}{\sqrt{\ex}}+\frac{N\ep}{2(\ex)^{3/2}}-\frac{N(E^{x'})^{2}}{8\ep(\ex)^{3/2}}+\frac{NE^{x''}}{2\ep\sqrt{\ex}} \nonumber \\
&&-\frac{NE^{\varphi'}E^{x'}}{2(\ep)^{2}\sqrt{\ex}}+\frac{N''\sqrt{\ex}}{\ep}+\frac{N'E^{x'}}{2\ep\sqrt{\ex}}-\frac{N'E^{\varphi'}\sqrt{\ex}}{(\ep)^{2}}-6\pi G\frac{N\ep(p^{x})^{2}}{(\ex)^{5/2}} \\
\label{ax dot}
\dot{A}_{x} &=& -\frac{\beta'}{4\pi}+\frac{N\ep p^{x}}{(\ex)^{3/2}}+N^{x'}A_{x}+N^{x}A_{x}' \\
\label{px dot}
\dot{p}^{x} &=& N^{x}p^{x'}
\eea

We have six equations of motion and three constraints to solve for six dynamical variables and three Lagrange multipliers. Now $C_{2}[\beta]=0$ implies that $p^{x}$ is independent of $x$. Equation \eqref{px dot} therefore implies that it is also independent of time $t$ thus implying that $p^{x}$ is a constant. We now choose the gauge $N^{x}=0$ and $E^{x}=x^{2}$ and look for static solution for the above set of equations. Note that the choice for $E^{x}$ in conjunction with \eqref{sphsymm adm metric} implies that $x$ refers to the radius of spherical sections. With these choices, \eqref{ex dot} and \eqref{ephi dot} imply that for non-zero $N$, $\kx=0=\kp$. This then implies that the diffeomorphism constraint is already satisfied. We are thus left with the Hamiltonian constraint and three equations of motion to solve for $N$, $\ep$ and $\beta$. We will thus have a non-trivial consistency condition telling us whether our gauge choice or the assumption of static solution is consistent or not. We find that a consistent solution is given by
\bea \label{classical sol}
\ep &=& x\left(1-\frac{2GM}{x}+\frac{GQ^{2}}{x^{2}}\right)^{-1/2} \\
N &=& \left(1-\frac{2GM}{x}+\frac{GQ^{2}}{x^{2}}\right)^{1/2}.
\eea
This determines $p^{x}=Q/\sqrt{4\pi}$ which we identify as the electric charge and $\beta=-\sqrt{4\pi}Q/x$ which is the electrostatic potential. Finally, substituting the solution \eqref{classical sol} in \eqref{sphsymm adm metric}, we have
\be \label{classical rn sol}
\md s^{2}=-\left(1-\frac{2GM}{x}+\frac{GQ^{2}}{x^{2}}\right)\md t^{2}+\left(1-\frac{2GM}{x}+\frac{GQ^{2}}{x^{2}}\right)^{-1}\md x^{2}+x^{2}\md \Omega^{2},
\ee
which we recognize as the classical Reissner-Nordstr\"om black hole solution with mass $M$ and charge $Q$.

\section{Inverse triad corrections}
In this section, we work out the consequences of incorporating the inverse triad corrections in the Hamiltonian constraint. These corrections arise when loop quantizing the inverse powers of triad variables. In the quantum theory, it is the flux variables that appear in the constraints. These are obtained by integrating the triad variables over two-dimensional surfaces and their spectrum is discrete containing zero, and thus, they do not have direct inverses. However, by using techniques from the full theory, an operator equation for the inverse can be defined indirectly which reproduces the classical inverse when quantum effects can be neglected \cite{thiemannQSDI, thiemannQSDV}. We can incorporate this correction at the effective level by replacing $1/\ex\rightarrow\alpha(\ex)/\ex$. The correction function $\alpha(\ex)$ can be obtained by making use of the techniques referred to above. 

In principle, the correction should depend not only on the macroscopic scale as provided by $\ex$ but should take into account the underlying discreteness of the theory. This is similar to the method of lattice refinement as used in cosmological models \cite{martinlatticeref, martininhomogeneities}. For homogeneous or symmetry-reduced models, the underlying discreteness of the full theory is not explicitly visible along the symmetry directions. To remain as close to the full theory as possible, one then has to include such effects by hand. From the full theory one expects that with an increasing area size (as determined by $\ex$) the underlying discreteness should get refined (since on macroscopic scales the spacetime is smooth). If the underlying discreteness is imagined as some kind of a lattice (realized via the spin network states) then these considerations would imply that with increasing area the number of discrete plaquettes forming a given spherical section should increase. Thus, in effect, the number of plaquettes is a phase-space-dependent function, which we write as $\mathcal{N}(\ex)$. 

\begin{figure}
\begin{center}
\includegraphics{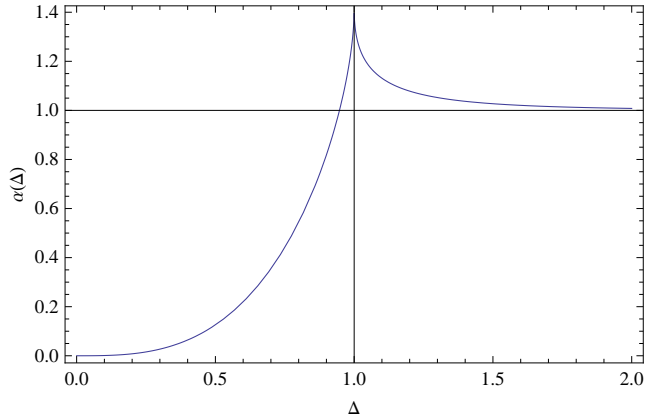}
\caption{\label{alphaDelta} The correction function $\alpha(\Delta)$ with $\Delta$ taken relative to
  $\Delta_*:=\sqrt{\gamma /2}\lP$.}
\end{center}
\end{figure}

In such a situation, the correct behavior of the correction function $\alpha$ is determined not by the scale $\ex$ but by the scale $\Delta\equiv\ex/\mathcal{N}(\ex)$ determining the plaquette size. For spherical symmetry, we find that 
the correction function is given by 
%(derivation given in \cite{LTB1})
\be \label{alpha}
\alpha(\Delta)=2\sqrt{\Delta}\frac{\sqrt{|\Delta+\gamma \lP^2/2|}-
\sqrt{|\Delta-\gamma \lP^2/2|}}{\gamma \lP^2}
\ee
(see \cite{LTB1} for a derivation of the above formula). Fig. \ref{alphaDelta} shows the behavior of the function and we can see that as $x\rightarrow0$, the classically singular behavior of $1/(\ex)^{1/2}$ is regularized and in the quantum domain instead of diverging it goes to zero. Thus, it is the plaquette size which controls the effectiveness of these corrections, and for small plaquette size, the corrections are large. From the figure we also see that the function $\alpha(\Delta)\rightarrow1$ quite fast once we are above the scale $\Delta_{*}=(\gamma/2)^{1/2}\lP$ and the corresponding quantum corrections are small. 
%  the effects of lattice refinement explicitly as far as inverse triad corrections are concerned since, as discussed in \cite{modified horizon}, its main effect (at least on short time scales) is to raise the scale where quantum effects become important by the factor $\mathcal{N}$. However this will not change the qualitative behavior of the solutions.  

We thus correct the classical Hamiltonian \eqref{ham constraint} to obtain
\begin{align} \label{inverse triad hamiltonian}
H^{Q}[N]=-\frac{1}{2G}\int \md x\, N\bigg[&\alpha|\ex|^{-\frac{1}{2}}\kp^2\ep+2\bar{\alpha}\kp\kx|\ex|^{\frac{1}{2}}+\alpha|\ex|^{-\frac{1}{2}}(1-\gp^2)\ep \notag\\
&+ 2\bar{\alpha}\gp'|\ex|^\frac{1}{2}-4\pi G\bar{\bar{\alpha}} \ep(p^{x})^{2}(\ex)^{-3/2} \bigg]\approx0.
\end{align}

In the above expression, for generality, different powers of $\ex$ are corrected by different $\alpha$'s. We would like to check whether demanding that the constraint algebra be first class even after incorporating quantum geometry corrections imposes any restrictions on the exact form of the correction functions $\bar{\alpha}$ and $\bar{\bar{\alpha}}$ (relative to $\alpha$). Note that formally $|E^{x}|^{1/2}$ can be written as $|\ex|/|\ex|^{1/2}$, and whereas the first factor will not pick up any quantum corrections, the second factor will have quantum corrections due to the inverse power of $\ex$ and thus to that extent $\bar{\alpha}$ is a reflection of quantization ambiguity \cite{QuantAmbiguity}. As already alluded to while discussing the classical constraints, in the matter part of the Hamiltonian we have $|\ex|^{-3/2}$ and this cannot be directly related to $|\ex|^{-1/2}$ and in principle, the correction $\bar{\bar{\alpha}}$ to this factor can be determined independently of $\alpha$. Thus, it would be interesting to see whether demanding a first class algebra restricts the function $\bar{\bar{\alpha}}$. 

Since the diffeomorphism constraint does not contain any inverse triad factors and because its action is directly represented on quantum states through group averaging, it is left unmodified. The Gauss constraint $C_{2}[\beta]$ is independent of triads, and therefore, it also retains its classical form. For this reason, the Poisson bracket between any two constraints of the pair $D[N^{x}]$ and $C_{2}[\beta]$ has its classical form. Similarly, $H^{Q}[N]$ Poisson commutes with $C_{2}[\beta]$ since the former is independent of $A_{x}$ whereas the latter depends only on $p^{x}$. We therefore only need to evaluate Poisson brackets $\{D[N^{x}],H^{Q}[N]\}$ and $\{H^{Q}[N],H^{Q}[M]\}$. After some straight forward but tedious algebra, one finds
\be \label{inverse triad constraint algebra}
\{D[N^{x}],H^{Q}[N]\}=H^{Q}[N'N^{x}], %+C_{2}[4\pi NN^{x}\bar{\bar{\alpha}}\ep p^{x}|\ex|^{-3/2}],
\ee
\be \label{quantum hh bracket}
\{H^{Q}[N],H^{Q}[M]\}=D[\bar{\alpha}^{2}|\ex|(\ep)^{-2}(NM'-N'M)],
\ee
and compared to the classical expression \eqref{classical hh bracket}, we note that there is an additional factor of $\bar{\alpha}$ on the r.h.s. Thus, even though the algebra is first class, compared to the classical algebra it has been deformed. In sections 3.1 and 3.2 we will explicitly show that while a classical algebra implies that under coordinate transformation the solution(s) of constraints and equations of motion map to other solution(s) implying that spacetime covariance holds, with a deformed algebra, the solution(s) of constraints and equations of motion do not map to other solution(s) implying that spacetime covariance is lost. 

Earlier in describing the form of the correction function $\alpha({\Delta})$ we had assumed that it depends only on $\ex$. A simplistic justification for this assumption can be provided by saying that only the inverse triad corrections due to $\ex$ are present, and therefore, correction function should also depend only on this variable. However, in the next section on holonomy corrections, we will see that such arguments are not always valid. In fact to start with, one can allow $\alpha$ to depend on $\ep$ as well. However, as discussed in \cite{LTB2}, $\ep$ dependence of $\alpha$ would lead to anomalous terms in the first Poisson bracket above. Demand of anomaly-free algebra is thus seen to restrict the form of the correction function.

Thus, we see that for arbitrary $\bar{\alpha}$ and $\bar{\bar{\alpha}}$, we have a first class algebra, with the only condition being that these be functions of $\ex$ alone. However, it is important to note that even though the algebra is first class, it does not have the classical form since now the bracket between two Hamiltonians depends on $\bar{\alpha}(\ex)$ (and $\bar{\alpha}(\ex)$, being a quantum correction, is not present in the classical theory). As has already been discussed in a previous paper \cite{modifiedhorizon} and as will be explicitly demonstrated in the following sub-sections, this implies that gauge transformations are no longer equivalent to coordinate transformations. The conventional spacetime picture is thus lost. 

On the other hand, we also note that since the condition of first class algebra does not restrict the form of $\bar{\alpha}$ (and because it is possibly related to quantization ambiguity), we can have $\bar{\alpha}=1$, in which case the Poisson bracket between two Hamiltonians has the classical form and the usual spacetime concepts hold. Thus, if we put the stronger condition that not only should the algebra be first class but should retain the classical form then we find that $\bar{\alpha}$ %(as also $\bar{\bar{\alpha}}$) 
is constrained to be unity. 
%Unlike $\bar{\alpha}$, however, $\bar{\bar{\alpha}}$ is not related to quantization ambiguity and thus in general the full constraint algebra would not have the classical form. 
In either case there is no restriction on the form of $\bar{\bar{\alpha}}$ thus corroborating the earlier observation that $\bar{\bar{\alpha}}$ correction is in principle unrelated to the $\alpha$ correction. In the present case we further find that it cannot be restricted even after imposing the stronger condition that the constraint algebra retains its classical form.

With the total Hamiltonian given by $H^{Q}_{T}=H^{Q}[N]+D[N^{x}]+C_{2}[\beta]$, the equations of motion are 
\bea \label{inverse triad exdot}
\dot{E}^{x} &=& N^{x}E^{x'}+2N\bar{\alpha}\kp\sqrt{\ex} \\
\label{inverse triad ephidot}
\dot{E}^{\varphi} &=& (N^{x}\ep)'+\frac{N\alpha\kp\ep}{\sqrt{\ex}}+N\bar{\alpha}\kx\sqrt{\ex} \\
\label{inverse triad kphidot}
\dot{K}_{\varphi} &=& N^{x}\kp'-\frac{\alpha N\kp^{2}}{2\sqrt{\ex}}-\frac{N\alpha}{2\sqrt{\ex}}-\frac{N\alpha(E^{x'})^{2}}{8(\ep)^{2}\sqrt{\ex}}+\frac{N\bar{\alpha}'E^{x'}\sqrt{\ex}}{2(\ep)^{2}}+\frac{N'\bar{\alpha}E^{x'}\sqrt{\ex}}{2(\ep)^{2}} \nonumber \\
&&+\frac{N\bar{\alpha}(E^{x'})^{2}}{4(\ep)^{2}\sqrt{\ex}}+2\pi G\frac{N\bar{\bar{\alpha}}(p^{x})^{2}}{(\ex)^{3/2}} \\
\label{inverse triad kxdot}
\dot{K}_{x} &=& (N^{x}\kx)'+\frac{N\alpha\kp^{2}\ep}{2(\ex)^{3/2}}-\frac{N\kp^{2}\ep}{\sqrt{\ex}}\frac{\delta\alpha}{\delta\ex}-\frac{N\bar{\alpha}\kp\kx}{\sqrt{\ex}}-2N\kp\kx\sqrt{\ex}\frac{\delta\bar{\alpha}}{\delta\ex} \nonumber \\
&&+\frac{N\alpha\ep}{2(\ex)^{3/2}}-\frac{N\ep}{\sqrt{\ex}}\frac{\delta\alpha}{\delta\ex}+\frac{N\alpha(E^{x'})^{2}}{8\ep(\ex)^{3/2}}-\frac{N(E^{x'})^{2}}{4\ep\sqrt{\ex}}\frac{\delta\alpha}{\delta\ex}-\frac{N'\alpha E^{x'}}{2\ep\sqrt{\ex}}-\frac{N\alpha E^{x''}}{2\ep\sqrt{\ex}} \nonumber \\
&&+\frac{N\alpha E^{\varphi'}E^{x'}}{2(\ep)^{2}\sqrt{\ex}}+\frac{N''\bar{\alpha}\sqrt{\ex}}{\ep}-\frac{N'\bar{\alpha}E^{\varphi'}\sqrt{\ex}}{(\ep)^{2}}+\frac{2N'E^{x'}\sqrt{\ex}}{\ep}\frac{\delta\bar{\alpha}}{\delta\ex}-\frac{2NE^{\varphi'}E^{x'}\sqrt{\ex}}{(\ep)^{2}}\frac{\delta\bar{\alpha}}{\delta\ex} \nonumber \\
&&-\frac{N\bar{\alpha}(E^{x'})^{2}}{4\ep(\ex)^{3/2}}+\frac{N(E^{x'})^{2}}{\ep\sqrt{\ex}}\frac{\delta\bar{\alpha}}{\delta\ex}+\frac{N'\bar{\alpha}E^{x'}}{\ep\sqrt{\ex}}+\frac{N\bar{\alpha}E^{x''}}{\ep\sqrt{\ex}}-\frac{N\bar{\alpha}E^{\varphi'}E^{x'}}{(\ep)^{2}\sqrt{\ex}}+\frac{NE^{x''}\sqrt{\ex}}{\ep}\frac{\delta\bar{\alpha}}{\delta\ex} \nonumber \\
&&+\frac{N\bar{\alpha}''\sqrt{\ex}}{\ep}-6\pi G\frac{N\bar{\bar{\alpha}}\ep(p^{x})^{2}}{(\ex)^{5/2}}+4\pi G\frac{N\ep(p^{x})^{2}}{(\ex)^{3/2}}\frac{\delta\bar{\bar{\alpha}}}{\delta\ex} \\
\label{inverse triad axdot}
\dot{A}_{x} &=& -\frac{\beta'}{4\pi}+\frac{N\bar{\bar{\alpha}}\ep p^{x}}{(\ex)^{3/2}}+N^{x'}A_{x}+N^{x}A_{x}' \\
\label{inverse triad pxdot}
\dot{p}^{x} &=& N^{x}p^{x'}
\eea

For this set of equations of motion and constraints, we now verify that a static solution is possible for arbitrary $\bar{\alpha}$ and $\bar{\bar{\alpha}}$. As before we have six equations of motion and three constraints and nine variables to solve for. As for the classical case, the constraint $C_{2}[\beta]=0$ implies that $p^{x}$ is independent of $x$ and then  \eqref{inverse triad pxdot} implies that $p^{x}$ is a constant (to be identified with the electric charge). Also note that \eqref{inverse triad axdot}, which determines $\beta$ in terms of $N$, $\ep$ and $\ex$ is decoupled from the rest of the equations. For a static solution, $A_{x}$ is independent of time and its dependence on $x$ is arbitrary which is fine since it just determines the electromagnetic gauge. For the static gauge we have $N^{x}=0$ and we choose $\ex=x^{2}$. Then, \eqref{inverse triad exdot} and \eqref{inverse triad ephidot} imply $\kp=0=\kx$. This implies that the diffeomorphism constraint is identically satisfied. We are thus left with the Hamiltonian constraint \eqref{inverse triad hamiltonian} and two equations of motion \eqref{inverse triad kphidot} and \eqref{inverse triad kxdot} to solve for two variables $N$ and $\ep$ thus implying a non-trivial consistency condition. 

With all this, the Hamiltonian constraint \eqref{inverse triad hamiltonian}, after substituting for $\Gamma_{\phi}$ and its derivative, simplifies to 
\be 
\frac{\alpha\ep}{x}-\frac{\alpha x}{\ep}-\frac{2\bar{\alpha}x}{\ep}+\frac{2\bar{\alpha}x^{2}E^{\varphi'}}{(E^{\varphi})^{2}}-\frac{4\pi G(p^{x})^{2}\bar{\bar{\alpha}}\ep}{x^{3}}=0,
\ee
implying the following differential equation for $\ep$:
\be \label{inverse triad ephiprime}
2\bar{\alpha}x^{5}E^{\varphi'}+\alpha x^{2}(\ep)^{3}-\alpha x^{4}\ep-2\bar{\alpha}x^{4}\ep-4\pi G(p^{x})^{2}\bar{\bar{\alpha}}(\ep)^{3}=0.
\ee
Similarly \eqref{inverse triad kphidot} can be simplified to give a differential equation for $N$
\be \label{inverse triad nprime}
2\bar{\alpha}x^{5}N'-\alpha x^{2}(\ep)^{2}N-\alpha x^{4}N+2\bar{\alpha}'x^{5}N+2\bar{\alpha}x^{4}N-4\pi G(p^{x})^{2}\bar{\bar{\alpha}}(\ep)^{2}N=0.
\ee
Use of these two equations along with the gauge choice and the solution $\kp=0=\kx$ shows that the r.h.s. of \eqref{inverse triad kxdot} is identically equal to zero, that is $\dot{\kx}=0$, thus proving that a static solution is possible for arbitrary $\bar{\alpha}$ and $\bar{\bar{\alpha}}$.

In the next two subsections we will discuss two cases: (I) $\bar{\alpha}=1$ where corrected Hamiltonian $H^{Q}$ results in modified dynamics but, with the constraint algebra being classical, the spacetime properties are unmodified and (II) $\bar{\alpha}=\alpha$ with a correspondingly modified constraint algebra implying that not only the dynamics but also the spacetime properties are modified. In this case, we will also try and see whether, through some suitable modifications, one can salvage conventional spacetime notions. 

At this stage we also have to decide on what form to choose for $\bar{\bar{\alpha}}$. As explained earlier, the form of $\alpha$ is determined by using techniques of \cite{thiemannQSDI, thiemannQSDV}. The method works to give quantum inverse for $(\ex)^{-p},\, 0<p<1$ whereas in the present case we require the inverse operator for $(\ex)^{-3/2}$ where this method does not work. Since $(\ex)^{-3/2}=((\ex)^{-1/2})^{3}$, we make the obvious choice $\bar{\bar{\alpha}}=\alpha^{3}$. Again there exist different possibilities for the form of $\bar{\bar{\alpha}}$ depending on how we view the corresponding classical expression, the above choice being just one example. There are thus quantum ambiguities present here which will modify the exact quantitative results. However, we expect that the qualitative behavior of the solution would not depend on such details.

\subsection{Case I: $\bar{\alpha}=1$}
We are interested in static solutions which, with a modified Hamiltonian, will imply quantum corrections to the Reissner-Nordstr\"om black holes. At the outset we mention that with only one type of quantum gravity correction included, whatever result we obtain cannot be considered as giving a complete picture, specially in the deep quantum regime where other effects like the holonomy corrections (to be dealt with in the next section) or quantum backreaction effects (not studied here) might become dominant. The present analysis should be seen as an exploration of the kind of quantum effects that result from different corrections and which correction(s) dominate in different regimes. At the present level of development, where not much is known about the quantum behavior in the full theory, such models can provide intuition for further developments.

With the gauge choice $N^{x}=0$ and $\ex=x^{2}$ already made, we need to solve equations \eqref{inverse triad ephiprime}, \eqref{inverse triad nprime} and \eqref{inverse triad axdot} to determine $\ep$, $N$ and $\beta$.  Furthermore, as mentioned in the beginning of the section, we need to include the effects of lattice refinement in our solutions such that the correction function $\alpha\equiv\alpha(\ex,\mathcal{N})$. In this sub-section we will consider two cases -- (i) $\mathcal{N}=\rm const$ corresponding to constant patch number and (ii) $\mathcal{N}\propto x^{p}$ corresponding to non-constant patch number. 

With $\bar{\alpha}=1$ and $\bar{\bar{\alpha}}=\alpha^{3}$, equations \eqref{inverse triad ephiprime} and \eqref{inverse triad nprime} become
\bea \label{alphabar1}
2x^{5}E^{\varphi'}-2x^{4}\ep-\alpha x^{4}\ep+\alpha x^{2}(\ep)^{3}-4\pi G\alpha^{3}(p^{x})^{2}(\ep)^{3} &=& 0, \label{eom ephi inverse triad 1} \\
2x^{5}N'-\alpha x^{2}(\ep)^{2}N-\alpha x^{4}N+2x^{4}N+4\pi G\alpha^{3}(p^{x})^{2}(\ep)^{2}N &=& 0. \label{eom n inverse triad 1}
\eea

To solve these equations we first note that classically $\ep_{c}=x/(1-2GM/x+GQ^{2}/x^{2})^{1/2}$ and $p^{x}=Q/\sqrt{4\pi}$. Also for this class of correction, the modified Schwarzschild solution as found in \cite{modifiedhorizon} gives $\ep_{s}=x/(1-2GMf_{\alpha}(x)/x)^{1/2}$, where the function $f_{\alpha}(x)$ is a solution of the equation $xf_{\alpha}'=(1-\alpha)f_{\alpha}$ with the subscript in $f_{\alpha}$ signifying the dependence of the functional form of $f$ on the refinement scheme. Since we know that for $Q=0$ our solution for $\ep$ should go over to that for modified Schwarzschild and that in the absence of quantum corrections it should go over to the classical solution, we make the ansatz $\ep=x/(1-2GMf_{\alpha}(x)/x+GQ^{2}b_{\alpha}(x)/x^{2})^{1/2}$ and substitute it in \eqref{alphabar1} to obtain an equation for $b_{\alpha}(x)$:
\be \label{eq for b}
xb_{\alpha}'-2b_{\alpha}+\alpha b_{\alpha}+\alpha^{3}=0.
\ee
Before solving this equation we find the form of the lapse function $N$ as well. We again make an ansatz which is motivated classically and also from the inverse triad corrected Schwarzschild solution and write $N=d_{\alpha}(x)(1-2GMf_{\alpha}(x)/x+GQ^{2}b_{\alpha}(x)/x^{2})^{1/2}/f_{\alpha}(x)$ with $d_{\alpha}(x)$ to be solved for. When used in \eqref{eom n inverse triad 1} one finds $d_{\alpha}(x)$ to be a constant which we choose to be unity to get the correct classical limit. Comparing with \eqref{sphsymm adm metric} the metric for this guage is then given by
\be \label{schwarzschild metric inverse triad 1}
\md s^{2}=-\frac{1}{f_{\alpha}^{2}}\left(1-\frac{2GMf_{\alpha}}{x}+\frac{GQ^{2}b_{\alpha}}{x^{2}}\right)\md t^{2}+\left(1-\frac{2GMf_{\alpha}}{x}+\frac{GQ^{2}b_{\alpha}}{x^{2}}\right)^{-1}\md x^{2}+x^{2}\md \Omega^{2}.
\ee
Although the theory based on which the above solution has been obtained (at an effective level) is non-perturbative, the form of the solution, where by the function $f_{\alpha}$ multiplies the mass $M$ and the function $b_{\alpha}$ multiplies the charge $Q$, suggests to draw an analogy with the renormalization concept familiar from perturbative quantum field theory and to think of these as the mass renormalization and charge renormalization, respectively, due to quantum gravitational effects \footnote{Terminology suggested by Romesh Kaul when shown some of the solutions.} (the renormalization concept itself being a physical effect independent of whether the underlying theory is perturbative or non-perturbative as exemplified by the renormalization group method). Similarly the metric coefficient $g_{tt}$ is corrected by $f_{\alpha}^{-2}$ and is akin to wave function renormalization. We further note that at the present effective level, Newton's constant $G$ does not seem to be renormalized due to quantum gravity effects. Using the expressions for $\ep$ and $N$ in \eqref{inverse triad axdot} we get the electrostatic potential
\be
\beta=\sqrt{4\pi}Q\int \md x\frac{\alpha^{3}}{f_{\alpha}x^{2}}\,,
\ee
and the factor $\alpha^{3}/f_{\alpha}$ contributes to the renormalization of the electrostatic potential. It is interesting to note that the renormalization of the electrostatic potential has different factor compared to that for electric charge.

With only the Hamiltonian constraint modified and the constraint algebra retaining its classical form, we expect that solutions of equations of motion and constraints will be mapped to other solutions under general coordinate transformation. As in \cite{modifiedhorizon}, we verify this by constructing the Painlev\'e-Gullstrand analogue of the above metric. Since the metric in \eqref{schwarzschild metric inverse triad 1} is time independent, $\xi_{t}=(1,0,0,0)$ is a Killing vector, and therefore, we can choose $g_{\alpha\beta}u^{\alpha}\xi^{\beta}_{t}=-1$, where $u^{\alpha}=\md x^{\alpha}/\md T$ is the four-velocity of the observer falling freely from infinity (starting at rest) in the Reissner-Nordstr\"om metric \eqref{schwarzschild metric inverse triad 1}. $T$ is the proper time along this trajectory and serves as the time coordinate for the Painlev\'e-Gullstrand metric. We also have $g_{\alpha\beta}u^{\alpha}u^{\beta}=-1$. Using these two equations we can solve for $u^{\alpha}$ which, with $\md T=-u_{\alpha}\md x^{\alpha}$, leads to
\be
\md T=\md t+\left(1-\frac{2GMf_{\alpha}}{x}+\frac{GQ^{2}b_{\alpha}}{x^{2}}\right)^{-1}\left(f_{\alpha}^{2}-1+\frac{2GMf_{\alpha}}{x}-\frac{GQ^{2}b_{\alpha}}{x^{2}}\right)^{1/2}\md x.
\ee
This when solved for $\md t$ and used in \eqref{schwarzschild metric inverse triad 1} gives the Painlev\'e-Gullstrand metric
\be \label{painleve metric inverse triad 1}
{\rm d}s^{2}=-{\rm d}T^{2}+f_{\alpha}^{-2}\left(\md x+\sqrt{f_{\alpha}^{2}-1+\frac{2GMf_{\alpha}}{x}-\frac{GQ^{2}b_{\alpha}}{x^{2}}}{\rm d}T\right)^{2}+ x^{2}{\rm d}\Omega^{2}.
\ee
Comparing this with \eqref{sphsymm adm metric}, we have $N=1$, $N^{x}=(f_{\alpha}^{2}-1+2GMf_{\alpha}/x-GQ^{2}b_{\alpha}/x^{2})^{1/2}$, $\ep=x/f_{\alpha}$ and $\ex=x^{2}$ in the Painlev\'e-Gullstrand gauge. It is straightforward to check that this solution satisfies all the constraints and the equations of motion. This once again confirms that for unmodified constraint algebra even if the dynamics is modified due to the modified Hamiltonian, solutions of constraints are mapped to other solutions under general coordinate transformations.

Having verfied that the spacetime picture is valid in the present case, we can now analyze the metric in \eqref{schwarzschild metric inverse triad 1} for horizon properties. We find that as opposed to the classical case, the metric coefficients $g_{tt}\neq g_{xx}^{-1}$. However, since in $g_{tt}$ the factor $f_{\alpha}^{-2}\geq0$ (this being zero only at $x=0$), the location of the horizon is given by the solution of 
\be \label{horizon inverse triad 1}
1-\frac{2GMf_{\alpha}}{x}+\frac{GQ^{2}b_{\alpha}}{x^{2}}=0.
\ee
To explore the horizon properties, we need to consider the explicit form of functions $f_{\alpha}$ and $b_{\alpha}$ for different refinement schemes. 

\begin{figure}
\begin{minipage}{0.5\linewidth}
\centering
\includegraphics[scale=.87]{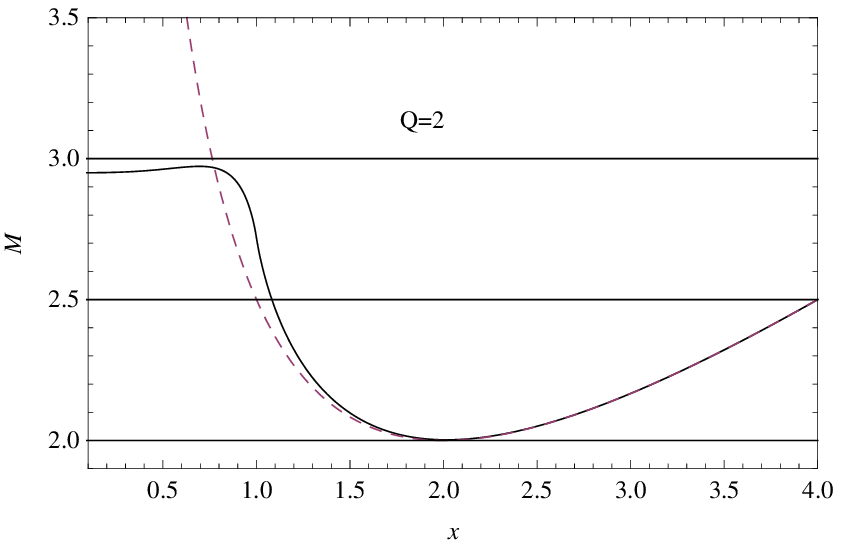}
\caption{\label{horizon cond q=2 inverse triad 1} The r.h.s. of \eqref{horizon curve inverse triad 1} (called the horizon curve in the text) for $Q=2$ with $x$ taken relative to
  $x_{*}:=\sqrt{\gamma\mathcal{N} /2}\lP$.}
\end{minipage}
\hspace{0.5cm}
\begin{minipage}{0.5\linewidth}
\centering
\includegraphics[scale=.72]{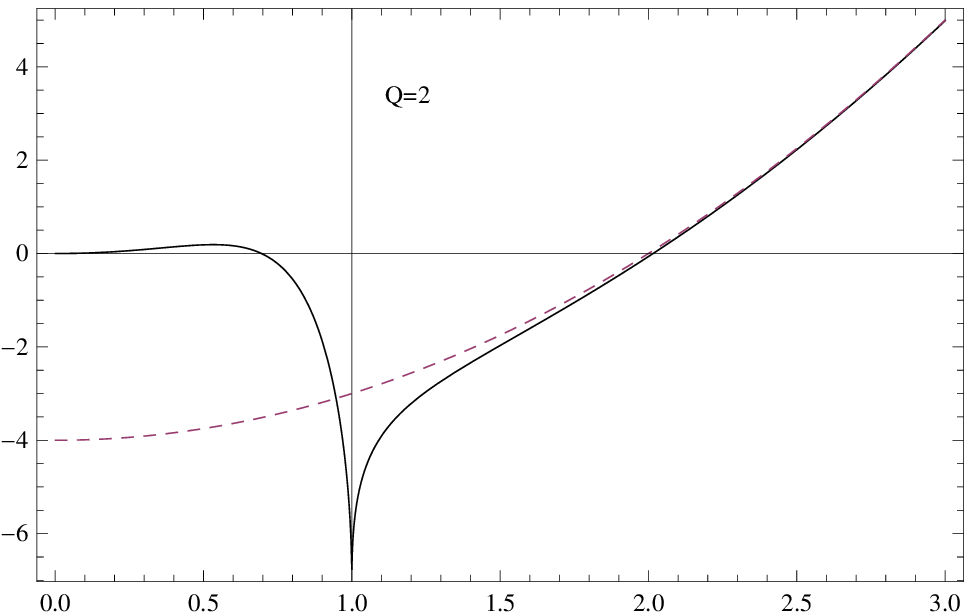}
\caption{\label{derivative horizon cond q=2 inverse triad 1} The factor $(x^{2}-\alpha^{2}Q^{2})$ (in the derivative of r.h.s. of \eqref{horizon curve inverse triad 1}) vs $x$ for $Q=2$ with $x$ taken relative to $x_{*}:=\sqrt{\gamma\mathcal{N} /2}\lP$.}
\end{minipage}
\end{figure}

\emph{Constant patch number}: For $\mathcal{N}=\text{constant}$, the solution for $f_{\alpha}$ as found in \cite{modifiedhorizon} is
\begin{figure}
\begin{minipage}{0.5\linewidth}
\centering
\includegraphics[scale=.8]{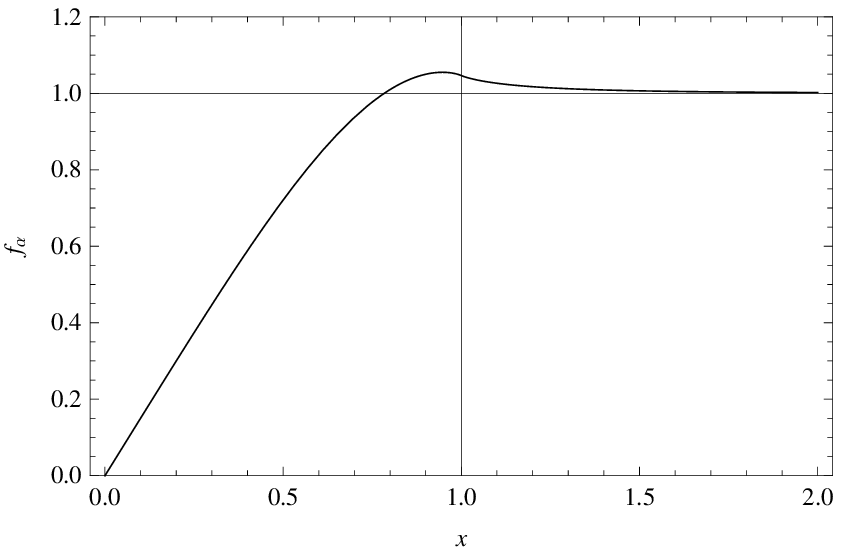}
\caption{\label{f plot} Function $f_{\alpha}(x)$ with $x$ taken relative to
  $x_{*}:=\sqrt{\gamma\mathcal{N} /2}\lP$.}
\end{minipage}
\hspace{0.5cm}
\begin{minipage}{0.5\linewidth}
\centering
\includegraphics[scale=.8]{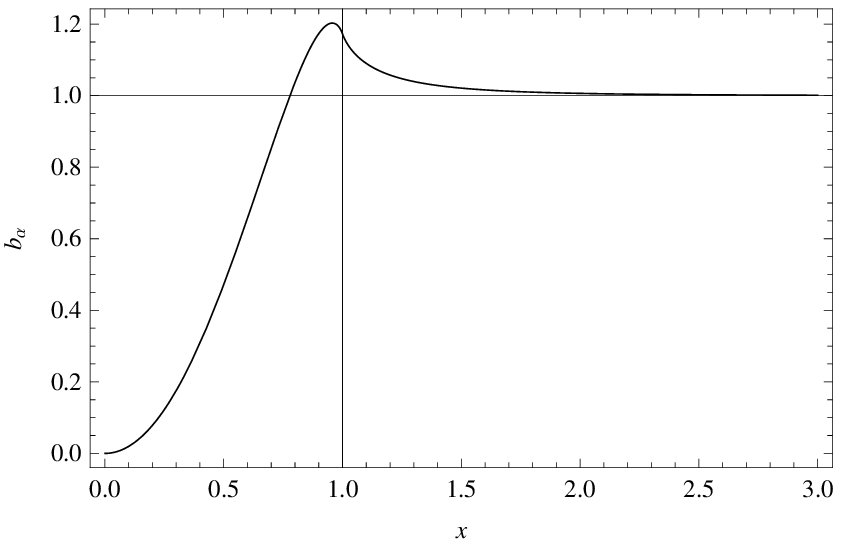}
\caption{\label{b plot} Function $b_{\alpha}(x)$ with $x$ taken relative to
  $x_{*}:=\sqrt{\gamma\mathcal{N} /2}\lP$.}
\end{minipage}
\end{figure} 
\bea 
f_{\alpha}(x) &=& \frac{2xe^{(1-\alpha)/2}}{\left(x+\sqrt{x^{2}-{\cal N}\gamma
    \lP^{2}/2}\right)^{1/2}\left(x+\sqrt{x^{2}+{\cal N}
\gamma \lP^{2}/2}\right)^{1/2}} \quad,\quad x^{2}>\mathcal{N}\gamma\lP^{2}/2 \\
\label{f for x less than planck scale}
f_{\alpha}(x) &=& \frac{2e^{-\pi/4}x e^{(1-\alpha(x))/2}e^{\frac{1}{2}\arctan
 \left(\sqrt{x^{2}/({\cal N}\gamma\lP^{2}/2-x^{2})}\right)}}{(
 {\cal N}\gamma\lP^{2}/2)^{1/4}
(x+\sqrt{x^{2}+{\cal N}\gamma\lP^{2}/2})^{1/2}} \quad,\quad x^{2}<\mathcal{N}\gamma\lP^{2}/2
\eea
which is plotted in Fig.~\ref{f plot}. Equation \eqref{eq for b} is difficult to solve analytically and we will therefore work with its numerical solution. For the purpose of numerics we choose $\mathcal{N}=1$ setting $G=1$ and measure $x$ in the units of $(\gamma\lP^{2}/2)^{1/2}$. Since in the classical limit $b(x)\rightarrow1$ and because the function $\alpha(\Delta)\rightarrow1$ very fast for $\Delta>\Delta_{*}$, we choose the boundary condition $b(100)=1$. The behavior of $b(x)$ is shown in Fig. \ref{b plot}. To analyze the horizon condition, we rewrite \eqref{horizon inverse triad 1} as (with $G=1$)
\be \label{horizon curve inverse triad 1}
M=\frac{x^{2}+Q^{2}b_{\alpha}}{2xf_{\alpha}}.
\ee
For the given values of $M$ and $Q$, the above equation will determine the location of the horizon. Since we do not have analytic solution for $b_{\alpha}$, we evaluate the horizon condition graphically. If we plot the r.h.s. of \eqref{horizon curve inverse triad 1} (which we call the horizon curve) as a function of $x$, then the points where the line $M=\text{constant}$ intersects this curve gives the location of the horizon. This is shown in Fig. \ref{horizon cond q=2 inverse triad 1} for $Q=2$, where for comparison we have also plotted the classical horizon condition $M=(x^{2}+Q^{2})/2x$ (shown with a dashed in the figure). 

We note that though the classical curve has only one extremum corresponding to extremal black holes, for the quantum corrected case we have two extremums. To reduce the dependence on the results of numerics, we extract more information directly from \eqref{horizon curve inverse triad 1}  by taking the $x$-derivative of the r.h.s. which works out to $\alpha(x^{2}-\alpha^{2}Q^{2})/2f_{\alpha}x^{2}$. Since $\alpha/2f_{\alpha}x^{2}>0$, we can extract the desired information from the factor $(x^{2}-\alpha^{2}Q^{2})$, which is plotted in Fig. \ref{derivative horizon cond q=2 inverse triad 1} and this again shows that there are two extremums. One of these extremums corresponds to the classical value, while the other is in deep quantum regime. 

Interestingly, the presence of non-classical extrema leads to the possibility of the black hole having three horizons as can be seen from Fig. \ref{derivative horizon cond q=2 inverse triad 1} where we find that the slope of the horizon curve on the left of the inner (non-classical) extremum is positive implying that for the choice $Q=2$, there are values of the mass of the black hole which can lead to three horizons (this feature is difficult to make out from Fig. \ref{horizon cond q=2 inverse triad 1} since the slope of the curve in this region is very small). 

Another very interesting feature is that there now seems to be only a finite range of mass, approximately $2<M<3$, for which there exist two (or possibly three) horizons. For $M\gtrsim3$ there exists only one horizon (corresponding to the outer horizon of classical black hole) and thus one of the distinguishing features of the Reissner-Nordstr\"om black hole is lost. Since in the deep quantum regime other effects, like holonomy corrections and quantum backreaction effects which are not considered here, can play a significant role, we do not analyze the implications of the presence of the extra horizon (and for the same reason we do not attempt to draw a Penrose diagram corresponding to this solution).

\begin{figure}
\begin{minipage}{0.5\linewidth}
\centering
\includegraphics[scale=.82]{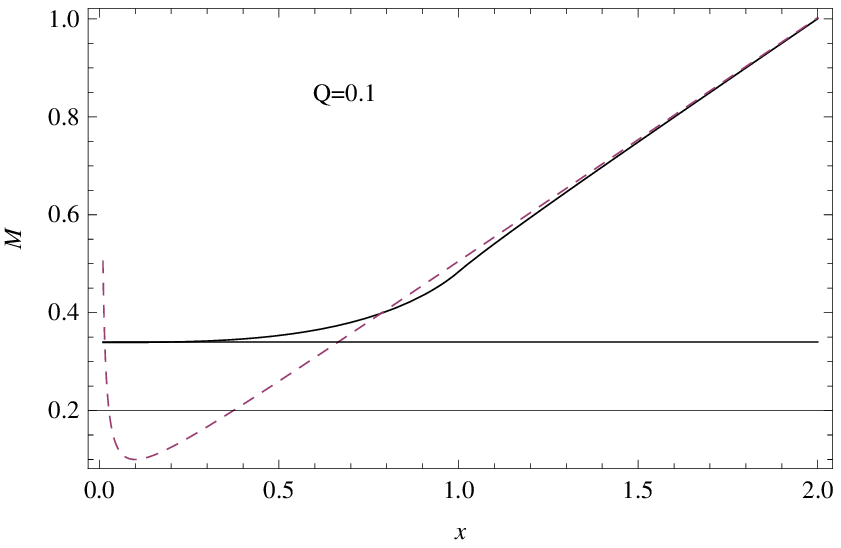}
\caption{\label{horizon cond q=0.1 inverse triad 1} The r.h.s. of \eqref{horizon curve inverse triad 1} (called the horizon curve in the text) for $Q=0.1$ with $x$ taken relative to  $x_{*}:=\sqrt{\gamma\mathcal{N} /2}\lP$.}
\end{minipage}
\hspace{0.5cm}
\begin{minipage}{0.5\linewidth}
\centering
\includegraphics[scale=.72]{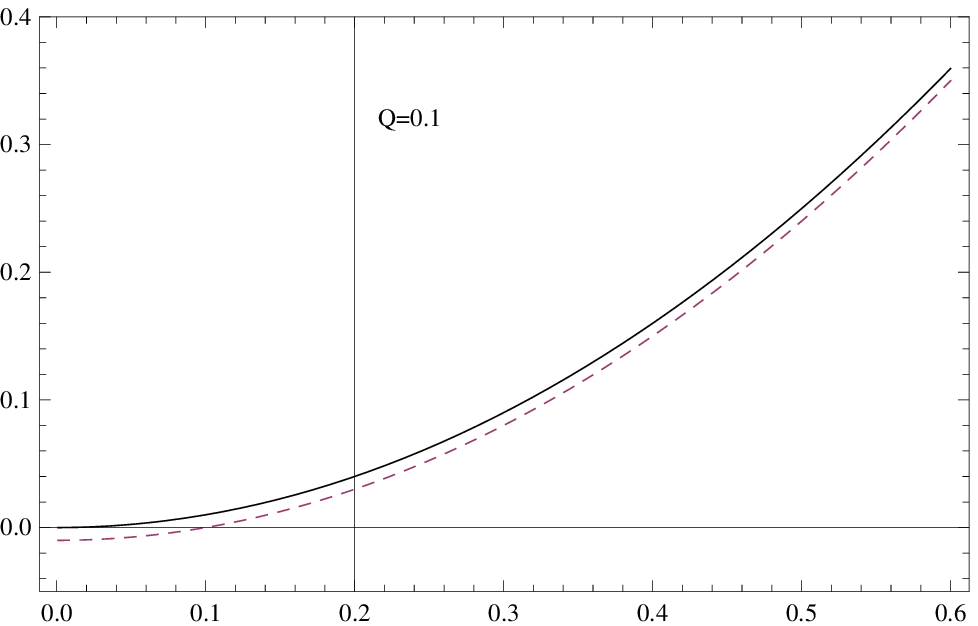}
\caption{\label{derivative horizon cond q=0.1 inverse triad 1} The factor $(x^{2}-\alpha^{2}Q^{2})$ (in the derivative of r.h.s. of \eqref{horizon curve inverse triad 1}) vs $x$ for $Q=0.1$ with $x$ taken relative to $x_{*}:=\sqrt{\gamma\mathcal{N} /2}\lP$.}
\end{minipage}
\end{figure}

To explore the dependence of the horizon properties on the value of the charge of the black hole, we now consider the case where the charge of the black hole is small, say $Q=0.1$. The relevant plots in Figs. \ref{horizon cond q=0.1 inverse triad 1} and \ref{derivative horizon cond q=0.1 inverse triad 1} now show different features and we find that for inverse triad corrected solution there is now only one horizon for any value of $M>0.34$ (in Planck units with $\gamma$ absorbed) and no horizon forms below this value. This is in agreement with the results in \cite{massthreshold, viqar critical, viqar mass gap, viqar modified GR, modifiedhorizon} where there is a mass threshold for the formation of horizon. 

Furthermore, there is no extremal solution possible in this case. For the classical black hole on the other hand, the extremal solution corresponds to $M=0.1$ and for mass greater than this value there are two horizons. For smaller values of charge $Q$, one is tending to the Schwarzschild limit of the Reissner-Nordstr\"om black hole and the mass threshold found above is comparable to the mass threshold for the formation of Schwarzschild black hole with inverse triad corrections \cite{modifiedhorizon}. 

To find the largest value of $Q$ for which these features persist, we note that for non-existence of extremal solution, the condition $(x^{2}-\alpha^{2}Q^{2})>0$ should hold for all $x$. For the given form of $\alpha$, this implies that for $Q<1/\sqrt{2}$ we have features similar to $Q=0.1$, while for $Q>1/\sqrt{2}$ one obtains the features discussed above for the case $Q=2$ except that for large enough values of $Q$, such that $(x^{2}-\alpha^{2}Q^{2})<0$ the inner extremum disappears and the slope to the left of the (outer) minimum in Fig.~\ref{horizon cond q=2 inverse triad 1} remains negative, and therefore, we do not get three horizons. However, there is still a mass threshold beyond which the inner horizon disappears.

In all the cases discussed above, one notes that the geometry for the quantum corrected case quickly merges with the classical geometry as one moves beyond the Planck scale and that all the interesting features correspond to deep quantum regime where, as already mentioned, role of other quantum effects needs to be taken into account to draw reliable conclusions. We also find that the singularity is not resolved with Ricci scalar diverging at $x=0$ (classically, Ricci scalar is zero and it is the Kretschmann scalar which diverges).

\emph{Non-constant patch number}: For a non-constant patch number we choose a power-law ansatz such that $\mathcal{N}\propto x^{p}$. This includes the special case where the patch size $\Delta=\ex/\mathcal{N}$ is a constant corresponding to $p=2$. In general, we can write $p=2+\epsilon$ with $\epsilon=0$ corresponding to the case with constant patch size. The correction function $\alpha$ in \eqref{alpha} takes the form
\begin{equation} \label{alpha with refinement}
\alpha=\frac{\sqrt{|1+a^{2}b^{2}x^{\epsilon}|}-\sqrt{|1-a^{2}b^{2}x^{\epsilon}|}}{a^{2}b^{2}x^{\epsilon}}\,,
\end{equation}
where $a^{2}\equiv \gamma\ell_{\rm P}^{2}/2$ and $b$ is a dimensionful constant needed to make $\mathcal{N}=b^{2}x^{2+\epsilon}$ dimensionless. As has been observed in \cite{modifiedhorizon}, only for $\epsilon<0$ we obtain physically reasonable solutions. For $\epsilon>0$, the patch size becomes smaller as we move to orbits with larger radius (and the correction function $\alpha$ tends to zero for large $x$ instead of tending to one), implying that quantum effects due to inverse triad corrections become dominant which is unphysical. We now have three parameters in the problem, i.e. $b$, $\epsilon$ and $Q$, and numerically it is difficult to explore the full parameter space. We looked at a few representative cases on which we now briefly comment. 

For $b=0.2$ ($a=1$ in the chosen units), $\epsilon=-1$ and $Q=2$, the behavior of the solution is very similar to the quantum corrected case with a constant patch number with $Q=2$. Similar behavior is observed when $b=1$, with other parameters remaining the same. Also for $Q=0.1$, the inner horizon seems to disappear which was also the case for constant patch number. This is consistent behavior since in this case the patch size is growing with increasing $x$ as is also the case when the patch number $\mathcal{N}$ is constant. On the other hand, when $\epsilon$ is close to zero, say $\epsilon=-0.1$, the behavior is very similar to a classical solution for different values of $b$ and $Q$. 

When $\epsilon=0$, the correction function $\alpha$ is a constant and the requirement that in the classical regime it should be close to one implies that $ab\ll1$ and it turns out that $\alpha$ is close to but greater than one. In this case, we can analytically solve \eqref{eq for b} for $b_{\alpha}$ (where we do not substitute the form \eqref{alpha with refinement} but use $\alpha$ itself remembering that it is a constant):
\be
b_{\alpha}=c_{1}x^{2-\alpha}-\frac{\alpha^{3}}{\alpha-2},
\ee
where $c_{1}$ is a constant. Since $\alpha$ is close to (and greater than) one, we find that imposing the desired limit $b_{\alpha}\rightarrow1$ for $x\rightarrow\infty$ forces one to choose $c_{1}=0=b$. For this case the function $f_{\alpha}$ is \cite{modifiedhorizon}
\be
f_{\alpha}(x)=c_{3}x^{1-\alpha},
\ee
and we see that $f_{\alpha}$ is very slowly decaying function of $x$ remaining almost constant for a large range of radial coordinate $x$. Thus, in general, both $f_{\alpha}$ and $b_{\alpha}$ do not have the expected large $x$ behavior of tending to unity. Both the functions however are very slowly varying functions of $x$ (for suitably chosen $c_{1}$ and $c_{3}$) and, as for the quantum corrected Schwarzschild in \cite{modifiedhorizon}, lead to a metric which in the asymptotic limit is conformally flat. %This looks reasonable since for $\epsilon=0$ the patch size is constant and therefore the behavior should be similar to the case when $\epsilon\approx0$ which was briefly discussed in the previous paragraph. 
  
%{\bf Check the validity of numerics by doing approximate analysis of the differential equation in asymptotic limit. What happens to extremal black holes, does some obvious change occurs for horizon and naked singularity? Check whether singularity occurs at $x=0$. Isolated horizons. Check how easy, difficult it would be to add magnetic charge.}

\subsection{Case II: $\bar{\alpha}=\alpha$}
We now put $\bar{\alpha}=\alpha$ everywhere in the set \eqref{inverse triad ephiprime} and \eqref{inverse triad nprime} (remember that already the gauge choice $N^{x}=0$ and $\ex=x^{2}$ has been made in writing these equations).  Proceeding as in the previous case, we make the classically motivated ansatz $\ep=x/(1-2GM/x+GQ^{2}g_{\alpha}(x)/x^{2})^{1/2}$ (for this version of inverse triad corrections, it turns out that there in no correction function in the mass term, see \cite{modifiedhorizon}). Substituting this in \eqref{inverse triad ephiprime} we get
\be \label{eq for g}
xg_{\alpha}'=g_{\alpha}-\alpha^{2}.
\ee
To solve for $N$ we make the ansatz $N=h_{\alpha}(x)(1-2GM/x+GQ^{2}g_{\alpha}(x)/x^{2})^{1/2}$ which on substituting in \eqref{inverse triad nprime} and making use of \eqref{eq for g} gives
\be \label{eq for h}
\frac{h'_{\alpha}}{h_{\alpha}}=-\frac{\alpha'}{\alpha}
\ee
with the solution $h_{\alpha}(x)=1/\alpha$ (where the constant of integration has been chosen to be 1). With this it turns out that
\be 
\beta=\sqrt{4\pi}Q\int \md x\frac{\alpha^{2}}{x^{2}}.
\ee
Compared to the classical solution there is an extra factor of $\alpha^{2}$ in the integrand which can be thought of as the renormalization of the electrostatic potential due to quantum gravity effects.

With the phase space solution known, the next task would be to write a spacetime metric corresponding to this solution. The obvious choice for the metric based on the solution of phase space variables would be
\be \label{non covariant metric schwarzschild}
\not\!{\rm d}s^{2}=-\frac{1}{\alpha^{2}}\left(1-\frac{2GM}{x}+\frac{GQ^{2}g_{\alpha}(x)}{x^{2}}\right)\md t^{2}+\left(1-\frac{2GM}{x}+\frac{GQ^{2}g_{\alpha}(x)}{x^{2}}\right)^{-1}\md x^{2}+x^{2}\md \Omega^{2}.
\ee
However, as has already been noticed in \cite{modifiedhorizon} in the context of vacuum Schwarzschild solution as well as for perturbative matter contributions, such a solution in not covariant (the slash in $\not\!{\rm d}s^{2}$ indicates that this is a formal construct and does not correspond to a covariant object). By this we mean that proceeding analogously to the previous sub-section if we perform a coordinate transformation on this metric to go over to the Painlev\'e-Gullstrand like coordinates, we find that the new metric does not satisfy the complete set of equations of motion and constraints. Specifically, following the standard procedure, one finds that in terms of the Painlev\'e-Gullstrand like coordinates the metric in \eqref{non covariant metric schwarzschild} becomes
\be \label{non covariant metric painleve}
\not\!{\rm d}s^{2}=-\md T^{2}+\frac{1}{\alpha^{2}}\left(\md x+\sqrt{\alpha^{2}-1+\frac{2GM}{x}-\frac{GQ^{2}g_{\alpha}(x)}{x^{2}}}\md T\right)^{2}+x^{2}\md \Omega^{2},
\ee
from which comparing with \eqref{sphsymm adm metric} we identify $N=1$, $\ep=x/\alpha$, $N^{x}=(\alpha^{2}-1+2GM/x-GQ^{2}g_{\alpha}(x)/x^{2})^{1/2}$ and $\ex=x^{2}$. 

Even though this form for the phase space variables (as also the metric in \eqref{non covariant metric painleve}) has the correct classical limit, these functions fail to satisfy the Hamiltonian constraint. Thus we see that for the case of modified constraint algebra, solution(s) of constraints (and of equations of motion) do not map to other solutions of constraints under general coordinate transformation. This has to be contrasted with the previous case for $\bar{\alpha}=1$, where the classical constraint algebra applied and hence solutions of constraints mapped to other solutions under a general coordinate transformation. 

With quantum corrections modifying the constraint algebra, we therefore seem to lose one of the most crucial aspects of the classical theory where gauge transformations in phase space correspond to coordinate transformations in spacetime and solutions of constraints are mapped to other solutions. In such a situation, where conventional spacetime notions lose their meaning, one cannot define concepts like metric, black holes and their horizons, which are used all the time in the classical theory. 

However, we also note that the constraint algebra is only deformed and has not become anomalous. Thus we expect that some notion of covariance should still survive. It is a non-trivial task to identify what this alternate notion could be, though whatever it be it certainly has to go over to the classical notion in the absence of quantum corrections. Since in the quantum theory a priori there is no spacetime but only phase space, it is possible that the mapping between the phase space and the spacetime that one is familiar with from the classical theory might only be a classical realization of a more general mapping. Below we provide one such alternative. However, at the outset we mention that the suggested mapping has not been obtained in a systematic manner and it is not clear how general it is. On the other hand, at the present stage of developments it is also useful to know various examples where similar methods work since they can provide hint for future developments. 

For the rest of this section, we need to distinguish between the lapse function occuring as part of the phase space $N_{\rm ps}$ (obtained from solving the constraints and equations of motion) from the lapse function $N_{\rm met}$ that will be used in the metric. We will show that although classically $N_{\rm ps}=N_{\rm met}$, for the case with inverse triad corrections where $\bar{\alpha}=\alpha$, a modified mapping $N_{\rm met}=\alpha N_{\rm ps}$ leads to a covariant metric. We have the following solution for the phase space variables: 
\be \label{phase space sol ver 2}
\ep=x\left(1-\frac{2GM}{x}+\frac{GQ^{2}g_{\alpha}(x)}{x^{2}}\right)^{-1/2} \quad,\quad N_{\rm ps}=\alpha^{-1}\left(1-\frac{2GM}{x}+\frac{GQ^{2}g_{\alpha}(x)}{x^{2}}\right)^{1/2}.
\ee
This would imply that under new mapping $N_{\rm met}$ is given by
\be \label{nmetric ver 2}
N_{\rm met}=\left(1-\frac{2GM}{x}+\frac{GQ^{2}g_{\alpha}(x)}{x^{2}}\right)^{1/2}.
\ee
With $\ex=x^{2}$ and $N^{x}=0$ we then have the `metric' corresponding to the classical Schwarzschild-like coordinates
\be \label{new metric ver 2 schwarzschild}
\md s^{2}=-\left(1-\frac{2GM}{x}+\frac{GQ^{2}g_{\alpha}(x)}{x^{2}}\right)\md t^{2}+\left(1-\frac{2GM}{x}+\frac{GQ^{2}g_{\alpha}(x)}{x^{2}}\right)^{-1}\md x^{2}+x^{2}\md \Omega^{2}.
\ee

Next we construct the Painlev\'e-Gullstrand analogue of the above metric. Following the procedure of the previous sub-section we find
\be
\md T=\md t+\left(1-\frac{2GM}{x}+\frac{GQ^{2}g_{\alpha}(x)}{x^{2}}\right)^{-1}\left(\frac{2GM}{x}-\frac{GQ^{2}g_{\alpha}(x)}{x^{2}}\right)^{1/2}\md x,
\ee
which when solved for $\md t$ and used in \eqref{new metric ver 2 schwarzschild} gives the Painlev\'e-Gullstrand metric:
\be \label{new metric ver 2 painleve}
\md s^{2}=-\md T^{2}+\left(\md x+\left(\frac{2GM}{x}-\frac{GQ^{2}g_{\alpha}(x)}{x^{2}}\right)^{1/2}\md T\right)^{2}+x^{2}\md \Omega^{2}.
\ee
From this we obtain $N_{\rm met}^{\rm PG}=1$, $\ep=x$, $\ex=x^{2}$ and $N^{x}=(2GM/x-GQ^{2}g_{\alpha}(x)/x^{2})^{1/2}$. To go to the phase space, we again need to use $N_{\rm met}=\alpha N_{\rm ps}$, which implies that $N_{\rm ps}^{\rm PG}=\alpha^{-1}$. It is now straightforward to check that this choice for the lapse function along with the form for $\ex$, $\ep$ and $N^{x}$ when used in equations \eqref{inverse triad exdot} and \eqref{inverse triad ephidot} determines $\kp=-N^{x}$ and $\kx=-N^{x'}$. It can now be verified that this solution satisfies the diffeomorphism constraint and the Hamiltonian constraint as well as the remaining equations of motion \eqref{inverse triad kphidot} and \eqref{inverse triad kxdot}. 

It is interesting to note that putting $Q=0$ in \eqref{new metric ver 2 schwarzschild} or \eqref{new metric ver 2 painleve} we get back the classical solution exactly and there is no trace of inverse triad corrections. Thus, if this modified mapping is used for vacuum solutions one would think that the covariance is recovered because one is effectively ignoring the quantum corrections using the new mapping. With $Q\neq0$, we clearly see that this is not the case since now there is a non-trivial quantum correction present in the form of function $g_{\alpha}(x)$. 

Here, we therefore have an example where, even though the constraint algebra is modified, one could still find a suitable notion of covariance. As required, the new mapping goes over to the classical mapping in the absence of quantum corrections since then $\alpha\rightarrow 1$. In the next section on holonomy corrections, where again the constraint algebra gets deformed, we will however see that in general it is much more difficult to identify an alternative notion of covariance and that probably we are lucky for the case of inverse triad corrections.

\begin{figure}
\begin{minipage}{0.5\linewidth}
\centering
\includegraphics[scale=.8]{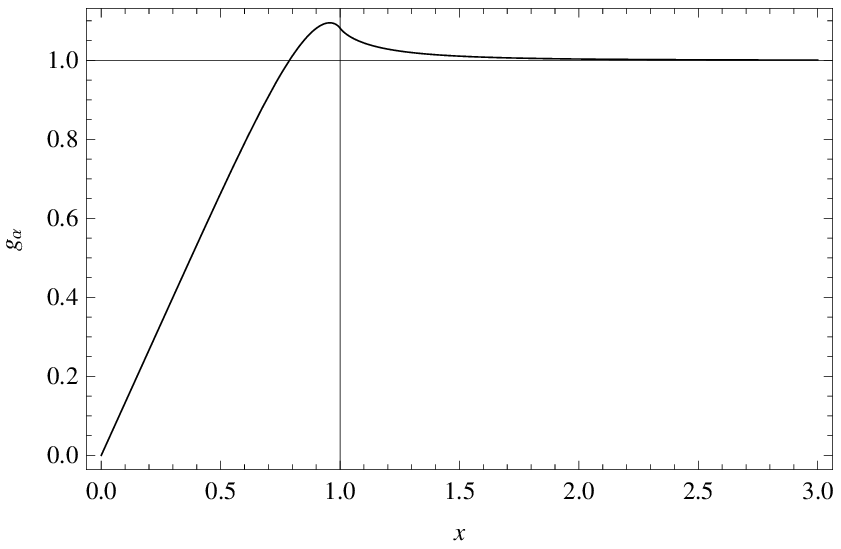}
\caption{\label{g plot} Function $g_{\alpha}(x)$ with $x$ taken relative to
  $x_{*}:=\sqrt{\gamma\mathcal{N} /2}\lP$.}
\end{minipage}
\hspace{0.5cm}
\begin{minipage}{0.5\linewidth}
\centering
\includegraphics[scale=.78]{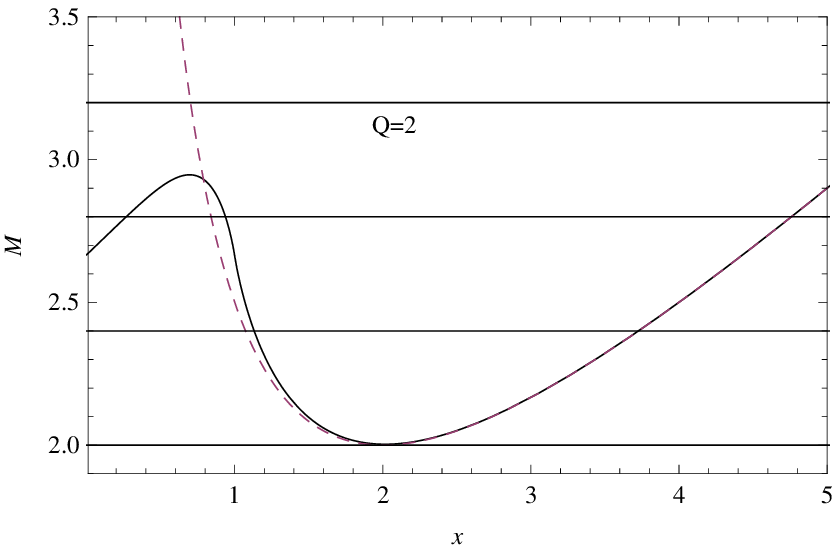}
\caption{\label{horizon cond q=2 inverse triad 2} The r.h.s. of \eqref{horizon curve inverse triad 1} (called the horizon curve in the text) for $Q=2$ with $x$ taken relative to
  $x_{*}:=\sqrt{\gamma\mathcal{N} /2}\lP$.}
\end{minipage}
\end{figure}

With a covariant solution in hand, it is now meaningful to look at the properties of the solution. From \eqref{new metric ver 2 schwarzschild} we find that compared to the previous case where $\bar{\alpha}=1$, in the current version of inverse triad corrections (and for the new mapping between phase space and spacetime) only the electric charge is renormalized and that there is no mass or wave function renormalization. However, as in that case, the gravitational constant is still not renormalized.  We now look at the horizon properties for this version of the inverse triad corrections also. From \eqref{new metric ver 2 schwarzschild}, we see that the horizon is given by the solution of 
\be  
1-\frac{2GM}{x}+\frac{GQ^{2}g_{\alpha}(x)}{x^{2}}=0,
\ee
which, as in the previous case, we write as (again putting $G=1$)
\be \label{horizon curve inverse triad 2}
M=\frac{x^{2}+Q^{2}g_{\alpha}}{2x}.
\ee
The behavior of function $g_{\alpha}$ is shown in Fig. \ref{g plot} and we see that it is very similar to $f_{\alpha}$ and $b_{\alpha}$. 

To analyze the horizon properties we proceed analogously to the previous case.   
We note that equation \eqref{horizon curve inverse triad 2} is similar to \eqref{horizon curve inverse triad 1} except that there is no $f_{\alpha}$ in the denominator. To explore the features of the horizon, as before we take derivative of the r.h.s. of the above equation with respect to $x$ which evaluates to $(x^{2}-\alpha^{2}Q^{2})/2x^{2}$. Barring the $2x^{2}$ in the denominator, this is exactly the expression used to discuss the properties of the horizon in the previous case (and shown in Fig. \ref{derivative horizon cond q=2 inverse triad 1}) and thus the conclusions drawn for the previous case where $\bar{\alpha}=1$ continue to hold in the present case also (though the exact values will differ). For this reason, we do not repeat the detailed analysis here except for showing a plot for the r.h.s. of \eqref{horizon curve inverse triad 2} for the case $Q=2$ in Fig.~\ref{horizon cond q=2 inverse triad 2}. 

In this figure, the inner extremum and the corresponding non-classical region to the left of it can be seen very clearly. We can also explicitly see the presence of the three horizons for a finite range of mass. As an example we have shown one such case where the horizontal line corresponding to $M=2.8$ intersects the curve at three different points. Again we note that the inner horizon disappears for mass greater than some finite value ($M\approx3$ in the case under consideration). Also for this version of inverse triad correction, the singularity is not resolved, though (un)like the (previous) classical case the Ricci scalar continues to be zero and it is the Kretschmann scalar which diverges.

%{\bf Approximate analytical solution for $g$(?), singularity? What happens to the concept of isolated horizon?}

\section{Holonomy effects}
We now move on to consider the effect of including holonomy corrections. These arise because in the quantum theory there is no representation for connection components as operators but only their holonomies have well defined operator analogues. Compared to inverse triad corrections, these corrections are of a different nature and are similar to (but not the same as) the higher derivative (higher curvature) terms in the effective theory. These corrections have not been analyzed much for spherically symmetric backgrounds though this has partly been done for marginally bound Lemaitre-Tolman-Bondi models in \cite{LTB1}. 

In the spherically symmetric background, there are two independent curvature components -- $\kp$ along the homogeneous angular directions and $\kx$ along the inhomogeneous radial direction. At an effective level that is being attempted here, one can incorporate holonomy effects by making a replacement $\kp\rightarrow \sin(\delta\kp)/\delta$ in the Hamiltonian, where $\delta$ refers to the path length used to calculate the holonomy and is a reflection of the scale of discreteness of the underlying quantum geometry. Since $\kp$ corresponds to homogeneous directions, the holonomies corresponding to this component are point holonomies familiar from loop quantum cosmology. 

This form of the function takes into account the fact that point holonomies correspond to replacing connection components by periodic functions in the Hamiltonian (the exact form of which will be found when fully working with the quantum theory rather then the effective approach being followed here) as also the fact that in the classical limit where the discreteness scale $\delta\rightarrow0$, we should recover the correct classical expression $\sin(\delta\kp)/\delta\rightarrow\kp$. 
%{\bf (Are $\delta$ and $\Delta$ similar objects and does $\delta\rightarrow0$ corresponds to $\Delta\rightarrow0$ of the inverse triad corrections? For inverse triad corrections it was seen that $\Delta\rightarrow0$ implies quantum effects dominate and therefore in classical regime corresponding to large $x$, $\Delta\not\!\rightarrow0$ while here we are saying that in the classical regime $\delta\rightarrow0$ and would be a contradiction if $\Delta$ and $\delta$ are similar objects)}. 

Incorporating the effect of $\kx$ holonomies is more non-trivial since they refer to the inhomogeneous radial direction. At the effective level, one can again attempt a replacement $\kx\rightarrow \sin(l\kx)/l$ where $l$ is the path length along the radial direction. However, because the radial direction is inhomogeneous, unlike the scale $\delta$ used for $\kp$ holonomy the scale $l\equiv l(x)$, there being a non-trivial dependence on the radial coordinate which needs to be handled carefully to ensure that the diffeomorphism invariance of the theory is not lost. Below we will only include the effects of the $\kp$ holonomy leaving the more complicated case of the $\kx$ holonomy for future work. 

As already mentioned, consistent inclusion of $\kp$ holonomies could be achieved in the marginal LTB model. However, a more systematic treatment for their inclusion has been given in \cite{juanthesis} working directly with constraints and obtaining the form of corrections by demanding that the first class nature of constraint algebra is preserved even after including these corrections. We will follow the same approach in the following.  

\subsection{Phase space independent holonomy corrections}
At the very beginning let us mention that though the title of the section seems to be a misnomer (since holonomies are dependent on the phase space variable $\kp$) what is meant is that the discreteness scale $\delta$ mentioned above is a constant independent of phase space variables.
As before we have three constraints in the theory and we will correct only the Hamiltonian constraint \eqref{ham constraint}, repeated here for convenience
\bea
H[N]&=&-\frac{1}{2G}\int \md x N|E^x|^{-\frac{1}{2}}(K_\varphi^2E^\varphi+2K_\varphi K_xE^x 
+ (1-\Gamma_\varphi^2)E^\varphi+2\Gamma_\varphi'E^x) \nonumber \\
&&+4\pi\int \md x\left(\frac{N\ep(p^{x})^{2}}{2(\ex)^{3/2}}\right).
\eea
We see that $\kp$ occurs at two places in the above expression and that these occurences are with different powers in the exponent. A priori there is no reason that the replacement $\kp\rightarrow\sin(\delta\kp)/\delta$ at both the places will give a consistent picture. Thus, to begin with, there is a two-fold functional degree of arbitrariness. To be general, therefore, we do not fix the form of the correction functions from the beginning and replace the quadratic $\kp^{2}$ in the first term by $f_{1}^{2}(\kp)$ and the linear $\kp$ in the second term by $f_{2}(\kp)$ (for the ease of notation we will not show the explicit depedence of $f_{1}, f_{2}$ on $\kp$ in the following). With this the Hamiltonian constraint becomes
\bea \label{ham constraint holonomy}
\bar{H}^{Q}[N]&=&-\frac{1}{2G}\int \md x N|E^x|^{-\frac{1}{2}}(f_{1}^{2}E^\varphi+2f_{2}K_xE^x 
+ (1-\Gamma_\varphi^2)E^\varphi+2\Gamma_\varphi'E^x) \nonumber \\
&&+4\pi\int \md x\left(\frac{N\ep(p^{x})^{2}}{2(\ex)^{3/2}}\right).
\eea

As for the case of inverse triad corrections, with only the Hamiltonian constraint receiving corrections, we only need to evaluate the Poisson brackets $\{D[N^{x}],\bar{H}^{Q}[N]\}$ and $\{\bar{H}^{Q}[N],\bar{H}^{Q}[M]\}$. Again a lengthy exercise in algebra gives
\be
\{D[N^{x}],\bar{H}^{Q}[N]\}=\bar{H}^{Q}[N'N^{x}], %+C_{2}[4\pi NN^{x}p^{x}\ep(\ex)^{-3/2}],
\ee
which is the same as the classical expression. The other bracket gives a more non-trivial result
\be \label{hh bracket holonomy}
\{\bar{H}^{Q}[N],\bar{H}^{Q}[M]\}=D[\frac{\partial f_{2}}{\partial\kp}\ex(\ep)^{-2}(NM'-N'M)]+\frac{1}{2G}\int \md z(NM'-N'M)\frac{E^{x'}}{\ep}\left(f_{2}-f_{1}\frac{\partial f_{1}}{\partial\kp}\right).
\ee
Clearly for a first class algebra the second term on the r.h.s. should vanish. This happens when the integrand in that term vanishes
\be \label{cond on holonomy corrections}
f_{2}-f_{1}\frac{\partial f_{1}}{\partial\kp}=0.
\ee
This gives a non-trivial condition on the form of the functions thus confirming that one cannot make the replacement $\kp\rightarrow\sin(\delta\kp)/\kp$ everywhere. To solve the above equation, we can choose a form for either $f_{1}$ or for $f_{2}$ and determine the form of the other function. Thus at the effective level there are several choices. However, it is interesting to note that demanding a first class algebra restricts the arbitrariness to just one functional degree of freedom. This we can fix by appealing to the full theory where point holonomies will appear as periodic functions. Since $f_{2}$ appears as a replacement for the linear $\kp$ in the Hamiltonian, it seems more appropriate to choose $f_{2}(\kp)=\sin(\delta\kp)/\delta$ and determine the form of $f_{1}$ rather then the other way round. With this choice for $f_{2}$ we find that the algebra remains first class if $f_{1}(\kp)=2\sin(\delta\kp/2)/\delta$. In general, $f_{2}=\sin(n\delta\kp)/n\delta$ implies $f_{1}=2\sin(n\delta\kp/2)/n\delta$. 

Thus, we find that like with inverse triad corrections a first class constraint algebra is possible even with holonomy corrections. However, unlike the case of inverse triad corrections, we note that in the present case we cannot obtain an unmodified algebra. To obtain an unmodified algebra, we would need to impose the condition $\partial f_{2}/\partial \kp=1$ (see \eqref{hh bracket holonomy}) which implies $f_{2}=\kp$ and \eqref{cond on holonomy corrections} then leads to $f_{1}=\kp$. Retaining classical algebra thus forces one to give up the holonomy corrections. 

In the full theory where all the corrections would be present simultaneously (including the holonomy corrections),  one would therefore expect that even if one could restrict the constraint algebra to be first class, it would be highly unlikely that it will retain its classical form. And it has already been observed that with a modified constraint algebra, there is no standard spacetime interpretation of the solutions to the constraints and the equations of motion.

\subsection{Phase space dependent holonomy corrections}
In the previous section we assumed that the holonomy corrections are independent of phase space. In principle, just like inverse triad corrections, these corrections should also be phase space dependent \cite{martinlatticeref, martininhomogeneities}. We have seen that $\delta$ refers to the scale of discreteness in the quantum theory. At the classical level, one would expect that this discreteness should tend to zero relative to a classical length scale. For spherically symmetric backgrounds under consideration, $\ex$ provides one such length scale, referring as it does to the area of spherical spatial sections for each choice of $x$. Thus one expects that the discreteness scale $\delta$, instead of being a constant should be a function of $\ex$ so that with increasing $\ex$, $\delta(\ex)$ tends towards zero. In loop quantum cosmology, the $\bar{\mu}$ scheme \cite{APS1} provides one realization of this dependence.

From the perspective of the full theory of quantum gravity, where the underlying quantum geometry is represented in terms of spin networks, the phase space dependence of discreteness scale $\delta$ in effect corresponds to the fact that under a Hamiltonian evolution, holonomies will create (and also destroy) new edges and vertices thus refining the graph underlying a given state of the theory. (One can get phase space dependent holonomy corrections even without lattice refinement if one combines them with the inverse triad corrections and demands a first class constraint algebra \cite{juanthesis}.)

For this section we therefore have $f_{1}\equiv f_{1}(\kp,\ex)$ and $f_{2}\equiv f_{2}(\kp,\ex)$. We again demand that with this choice of correction functions the constraint algebra be first class. Only the $\{\bar{H}^{Q}[N],\bar{H}^{Q}[M]\}$ is non-trivial and gives
\bea \label{hh bracket phase space holonomy}
\{\bar{H}^{Q}[N],\bar{H}^{Q}[M]\} &=& D[\frac{\partial f_{2}}{\partial\kp}\ex(\ep)^{-2}(NM'-N'M)] \nonumber \\
&&+\frac{1}{2G}\int \md z(NM'-N'M)\frac{E^{x'}}{\ep}\left(f_{2}-f_{1}\frac{\partial f_{1}}{\partial\kp}+2\ex\frac{\partial f_{2}}{\partial\ex}\right).
\eea
For a first class algebra we therefore require
\be \label{cond on phase space holonomy corrections}
f_{2}-f_{1}\frac{\partial f_{1}}{\partial\kp}+2\ex\frac{\partial f_{2}}{\partial\ex}=0,
\ee
and compared to \eqref{cond on holonomy corrections} there is an additional term. To solve this equation we choose $f_{2}(\kp,\ex)=\sin(\delta(\ex)\kp)/\delta(\ex)$. As before, a basic condition on such correction functions is that in the classical limit $f_{2}\rightarrow\kp$. Following the arguments in the beginning of this sub-section and noting that since $\kp$ is dimensionless so that $\delta(\ex)$ is also dimensionless, we choose $\delta(\ex)\thicksim(\gamma\lP^{2}/\ex)^{p}$, $p>0$. This choice automatically gives us the classical behavior for $\ex\gg\gamma\lP^{2}$. With this choice for $\delta(\ex)$ we have 
\be \label{f2}
f_{2}(\kp,\ex)=\left(\frac{\ex}{\gamma\lP^{2}}\right)^{p}\sin\left(\left(\frac{\gamma\lP^{2}}{\ex}\right)^{p}\kp\right),
\ee
and from the first term of \eqref{hh bracket phase space holonomy} we note that with a non-trivial $\kp$ dependence of $f_{2}$, undeformed constraint algebra is not possible. This form for $f_{2}$ when used in \eqref{cond on phase space holonomy corrections} gives two solutions (differing in an overall sign) of which the physically important one is
\bea \label{sol for f1}
f_{1,p} &=& (2)^{1/2}\biggl[-\left(\frac{\ex}{\gamma\lP^{2}}\right)^{p}\left((1+4p)\left(\frac{\ex}{\gamma\lP^{2}}\right)^{p}\cos\left(\left(\frac{\gamma\lP^{2}}{\ex}\right)^{p}\kp\right)+2p\kp\sin\left(\left(\frac{\gamma\lP^{2}}{\ex}\right)^{p}\kp\right)\right) \nonumber \\
&& +C_{p}(\ex)\biggr]^{1/2},
\eea
where %in the above equation 
$C_{p}(\ex)$ is an arbitrary function (since \eqref{cond on phase space holonomy corrections} involves partial derivative of $f_{1}$ only with respect to $\kp$, it does not completely fix the $\ex$ dependence). We can fix the form of $C_{p}(x)$ by demanding that $f_{1}\rightarrow\kp$ when $\gamma\lP^{2}/\ex\rightarrow0$. Doing a reduction consistently at order $\kp^{2}$, we find that $C_{p}(\ex)=(1+4p)(\ex)^{2p}/(\gamma\lP^{2})^{2p}$ gives the correct classical limit.

In the following, we will mostly work with an arbitrary value of $p$. However, certain well motivated considerations lead one to choose specific value(s) for $p$. In the previous section on inverse triad corrections,  the effects of lattice refinement were encoded in the number of patches $\mathcal{N}(\ex)$ which was seen to be related to the area of spherical sections. Similarly, in the present case, we know that $\delta(\ex)$ refers to the length of the curve on the spherically symmetric orbits along which the holonomy is evaluated. The lattice gets refined as one moves to spherical sections of larger radius and one possibility is that the number of lattice sites on a given orbit increases in proportion to the radius (corresponding to constant lattice size) and thus $\delta(\ex)$ itself should vary inversely as the radius implying that we choose $p=1/2$. For this value of $p$ \eqref{sol for f1} gives
\be \label{sol for f1 p equal half}
f_{1,1/2}=(2)^{1/2}\left[3\frac{\ex}{\gamma\lP^{2}}-3\frac{\ex}{\gamma\lP^{2}}\cos\left(\left(\frac{\gamma\lP^{2}}{\ex}\right)^{\frac{1}{2}}\kp\right)-\kp\left(\frac{\ex}{\gamma\lP^{2}}\right)^{\frac{1}{2}}\sin\left(\left(\frac{\gamma\lP^{2}}{\ex}\right)^{\frac{1}{2}}\kp\right)\right]^{1/2}.
\ee
 
At this point it is useful to note that \eqref{sol for f1} and \eqref{sol for f1 p equal half} are not periodic in $\kp$ (even though $f_{2}$ is) unless $p=0$. It is therefore not clear how such corrections can arise from the underlying theory where it is expected that point holonomies should appear as periodic functions \footnote{Author is thankful to Martin Bojowald for pointing this out.}. Demand of a first class constraint algebra for phase space dependent holonomy corrections thus seems to imply non-trivial corrections from the perspective of the full theory. It remains to be seen whether a more complete understanding of the off-shell closure of the constraint algebra in the full theory can give rise to such corrections.  

%As a side remark we note that if we demand that not only the constraint algebra be first class but that it be the same as the classical algebra then we have an extra condition on $f_{2}$, namely, $\partial f_{2}/\partial\kp=1$ coming from the first term on the right of \eqref{hh bracket phase space holonomy}. This can be integrated immediately to give $f_{2}=\kp+m(E^{x})$ where $m(E^{x})$ is an arbitrary function of $\ex$ which we can try to fix by appealing to the classical condition $f_{2}|_{\rm class}\rightarrow\kp$ which implies $m(\ex)|_{\rm class}\rightarrow0$. One obvious choice is $m(E^{x})=0$ but other non-trivial choice(s) also exist. For instance one can choose $m(\ex)=\alpha(\ex)-1$ where $\alpha(\ex)$ could be the same function as used in the previous section on inverse triad corrections or one can choose $m(\ex)=\sin(\gamma\lP^{2}/\ex)$ and for any of these choices one can solve \eqref{cond on phase space holonomy corrections} for $f_{1}$. Although both these choices give the correct classical behavior, none is motivated from the underlying theory and therefore do not represent phase space dependent holonomy corrections coming from loop quantum gravity. Thus with non-trivial \emph{holonomy} corrections only for $\kp$ the constraint algebra cannot be of the classical form.

We now look at the dynamics for this class of corrections. The Hamiltonian constraint is given by \eqref{ham constraint holonomy} (but we remember that $f_{1}$ and $f_{2}$ in that expression now depend on both $\kp$ and $\ex$) and the equations of motion are
\bea \label{holonomy exdot}
\dot{E}^{x} &=& N^{x}E^{x'}+2Nf_{2}\sqrt{\ex} \\
\label{holonomy ephidot}
\dot{E}^{\varphi} &=& (N^{x}\ep)'+\frac{Nf_{1}\ep}{\sqrt{\ex}}\frac{\partial f_{1}}{\partial\kp}+N\kx\sqrt{\ex}\frac{\partial f_{2}}{\partial\kp} \\
\label{holonomy kphidot}
\dot{K}_{\varphi} &=& N^{x}\kp'-\frac{Nf_{1}^{2}}{2\sqrt{\ex}}-\frac{N}{2\sqrt{\ex}}+\frac{N(E^{x'})^{2}}{8(\ep)^{2}\sqrt{\ex}}+\frac{N'E^{x'}\sqrt{\ex}}{2(\ep)^{2}}+2\pi G\frac{N(p^{x})^{2}}{(\ex)^{3/2}} \\
\label{holonomy kxdot}
\dot{K}_{x} &=& (N^{x}\kx)'+\frac{Nf_{1}^{2}\ep}{2(\ex)^{3/2}}-\frac{2Nf_{1}\ep}{\sqrt{\ex}}\frac{\partial f_{1}}{\partial\ex}-\frac{Nf_{2}\kx}{\sqrt{\ex}}-2N\kx\sqrt{\ex}\frac{\partial f_{2}}{\partial\ex}+\frac{N\ep}{2(\ex)^{3/2}} \nonumber \\
&&-\frac{N(E^{x'})^{2}}{8\ep(\ex)^{3/2}}+\frac{NE^{x''}}{2\ep\sqrt{\ex}}-\frac{NE^{\varphi'}E^{x'}}{2(\ep)^{2}\sqrt{\ex}}+\frac{N''\sqrt{\ex}}{\ep}+\frac{N'E^{x'}}{2\ep\sqrt{\ex}} \nonumber \\
&&-\frac{N'E^{\varphi'}\sqrt{\ex}}{(\ep)^{2}}-6\pi G\frac{N\ep(p^{x})^{2}}{(\ex)^{5/2}}
\eea
\bea
\label{holonomy axdot}
\dot{A}_{x} &=& -\frac{\beta'}{4\pi}+\frac{N\ep p^{x}}{(\ex)^{3/2}}+N^{x'}A_{x}+N^{x}A_{x}' \\
\label{holonomy pxdot}
\dot{p}^{x} &=& N^{x}p^{x'}
\eea

As in earlier cases, we try to solve the constraints and the equations of motion looking for static solutions with $N^{x}=0, \ex=x^{2}$ and as mentioned earlier, we do not fix a value for the refinement parameter $p$. However, in the present case, we find that for this gauge choice \eqref{holonomy exdot} gives $2Nf_{2}\sqrt{\ex}=0$. This implies that for non-zero $N$, $f_{2}=0$ which is possible only if $\kp=n\pi x^{2p}/(\gamma\lP^{2})^{2p}$, $n\in\mathbb I$. In such a situation even $f_{1}=0$ and except for the non-classical form of $\kp$ for $n\neq0$, nowhere else in the solution do we see the quantum effects. This is so because putting $f_{1}=0=f_{2}$ everywhere in the constraints and in equations of motion is equivalent to going back to the classical solution with $\kp=0=\kx$. This happens also for the case where we consider phase space independent holonomy corrections. 

We therefore find that choosing $N^{x}=0$ and demanding a static solution always gives the classical solution unless there is some non-trivial correction involving $\kx$ in the Hamiltonian (since \eqref{holonomy exdot} involves derivative of the Hamiltonian with respect to $\kx$). Thus $\kp$ holonomy corrections alone seem to preclude a static solution as well as an unmodified version of constraint algebra. However, a stationary solution might still be possible, which we look for next. 

To obtain time independent solution, we choose $\ex=x^{2}$ but keep the functional dependence of other phase space variables arbitrary. As before, $C_{2}[\beta]=0$ together with \eqref{holonomy pxdot} implies that $p^{x}$ is a constant which we again choose to be $Q/\sqrt{4\pi}$. 
%Also \eqref{holonomy axdot} determines $\beta$ in terms of $\ex$, $\ep$ and $p^{x}$. 
Equation \eqref{holonomy exdot} implies $N^{x}=-Nf_{2}$. Solving the diffeomorphism constraint gives $\kx=\kp'\ep/x$. Using these in \eqref{holonomy ephidot} then gives the relation $\ep=cx/N$, $c$ a constant. This is the (classically) expected relation between $\ep$ and $N$ (eg. the classical Reissner-Nordstr\"om solution) and we will choose $c=1$. Using these results in \eqref{holonomy kphidot} and in the Hamiltonian constraint \eqref{ham constraint holonomy} shows that these two equations are equivalent. Thus we are left with \eqref{holonomy kxdot}, \eqref{holonomy axdot} and the Hamiltonian constraint to solve for $\kp$, $N$ and $\beta$. The Hamiltonian constraint gives
\be 
\frac{f_{1}^{2}}{x}+2f_{2}\kp'=-\frac{1}{x}+\frac{3x}{(\ep)^{2}}-\frac{2x^{2}E^{\varphi'}}{(\ep)^{3}}+\frac{GQ^{2}}{x^{3}}.
\ee  
Replacing $\ep$ by $x/N$ this equation can be written as
\be \label{solving holonomy ham}
\frac{f_{1}^{2}}{x}+2f_{2}\kp'=-\frac{1}{x}+\frac{N^{2}}{x}+2NN'+\frac{GQ^{2}}{x^{3}}.
\ee
We note that the left hand side of this equation depends only on $\kp$ and the right side only on $N$. Next we write \eqref{holonomy kxdot} entirely in terms of $\kp$ and $N$ to obtain
\be 
-f_{2}'\kp'-f_{2}\kp''+\frac{f_{1}^{2}}{2x^{2}}-2f_{1}\frac{\partial f_{1}}{\partial\ex}-\frac{f_{2}\kp'}{x}-2x\kp'\frac{\partial f_{2}}{\partial\ex}=-\frac{1}{2x^{2}}+\frac{N^{2}}{2x^{2}}-\frac{NN'}{x}-NN''-(N')^{2}+\frac{3GQ^{2}}{2x^{4}}.
\ee
As for the Hamiltonian constraint, this equation separates into two parts - left hand side depending on $\kp$ and the right-hand side depending on N. It turns out that the right-hand side of the above equation is nothing but one-half of the $x$-derivative of the r.h.s. of \eqref{solving holonomy ham} and can therefore be replaced by one-half of the $x$-derivative of the l.h.s. of \eqref{solving holonomy ham} to obtain the equation entirely in terms of $\kp$. Finally, substituting for $f_{1}$ and $f_{2}$ from \eqref{sol for f1} and \eqref{f2} along with the required derivatives, one finds that the above equation is identically equal to zero, that is, $\dot{\kx}=0$ is identically satisfied. 

Thus, with the choice $\ex=x^{2}$ and the assumption of time independence, we have solved all the equations except for \eqref{solving holonomy ham} with $N$ and $\kp$ still undetermined (of course \eqref{holonomy axdot} is also there but it just determines $\beta$). We therefore have the freedom to arbitrarily choose one of the functions. We now present the solution to \eqref{solving holonomy ham} which classically would correspond to the Reissner-Nordstr\"om solution in \eqref{classical rn sol}. To obtain such a solution we make the choice
\be \label{holonomy n schwarzschild}
N=\left(1-\frac{2GM}{x}+\frac{GQ^{2}}{x^{2}}\right)^{1/2}.
\ee
This using $\ep=x/N$ implies
\be \label{holonomy ephi schwarzschild}
\ep=x\left(1-\frac{2GM}{x}+\frac{GQ^{2}}{x^{2}}\right)^{-1/2}.
\ee
Using this in \eqref{solving holonomy ham} we obtain
\be \label{holonomy kp eq schwarzschild}
\frac{f_{1}^{2}}{x}+2f_{2}\kp'=0,
\ee
which on substituting for $f_{1}$ and $f_{2}$ leads to
\bea
&&2(4p+1)\frac{x^{4p-1}}{(\gamma\lP^{2})^{2p}}-2(4p+1)\frac{x^{4p-1}}{(\gamma\lP^{2})^{2p}}\cos\left(\frac{(\gamma\lP^{2})^{p}}{x^{2p}}\kp\right)-4p\frac{x^{2p-1}}{(\gamma\lP^{2})^{p}}\kp\sin\left(\frac{(\gamma\lP^{2})^{p}}{x^{2p}}\kp\right) \nonumber \\
&&+2\frac{x^{2p}}{(\gamma\lP^{2})^{p}}\kp'\sin\left(\frac{(\gamma\lP^{2})^{p}}{x^{2p}}\kp\right)=0.
\eea
Making use of the trigonometric half-angle formulae, the above equation becomes
\be
(4p+1)\frac{x^{2p}}{(\gamma\lP^{2})^{p}}\sin\left(\frac{(\gamma\lP^{2})^{p}}{2x^{2p}}\kp\right)-2p\kp\cos\left(\frac{(\gamma\lP^{2})^{p}}{2x^{2p}}\kp\right)+x\kp'\cos\left(\frac{(\gamma\lP^{2})^{p}}{2x^{2p}}\kp\right)=0,
\ee
where in writing the above equation we have ignored the possible $\sin((\gamma\lP^{2})^{p}\kp/2x^{2p})=0$ solution since, as noted earlier, this corresponds to a trivial solution from the point of view of quantum dynamics. To solve this equation we make the ansatz $\kp=2x^{2p}\sin^{-1}(r)/(\gamma\lP^{2})^{p}$ and solve for $r$ to find $r=c/x^{(4p+1)/2}$, where $c$ is an arbitrary constant. This implies that
\be \label{holonomy kp schwarzschild}
\kp=\frac{2x^{2p}}{(\gamma\lP^{2})^{p}}\sin^{-1}\left(\frac{c}{x^{(4p+1)/2}}\right).
\ee 
To fix the constant $c$, we first note that it has dimension $[L^{(4p+1)/2}]$ and secondly that when quantum effects tend to zero (as characterized by $\gamma\lP^{2}$ tending to zero for a fixed $x\gg\sqrt{\gamma\lP^{2}}$),  $\kp$ should tend to the classical value of zero. Then, the obvious choice for $c$ that suggests itself is $c=(\gamma\lP^{2})^{(4p+1)/4}$. We thus have the complete solution with $N^{x}=-Nf_{2}$ after substituting for $N$, $f_{2}$ and $\kp$ evaluating to 
\be \label{holonomy nx schwarzschild}
N^{x}=-2\frac{(\gamma\lP^{2})^{1/4}}{x^{1/2}}\left(1-\frac{(\gamma\lP^{2})^{(4p+1)/2}}{x^{4p+1}}\right)^{1/2}\left(1-\frac{2GM}{x}+\frac{GQ^{2}}{x^{2}}\right)^{1/2}
\ee 
and we note that in the absence of quantum effects this correctly gives $N^{x}=0$. 

With the solution for the phase space variables in hand the obvious thing to do would be to write a metric corresponding to this solution. However, as seen in the last section for version II of inverse triad corrections, with modified constraint algebra conventional spacetime notions including the usual notion of metric do not make sense since the covariance property as encoded in the constraint algebra is modified. Nevertheless, there certainly is gauge invariance in the theory since the constraint algebra is not anomalous, it is just deformed. Thus, once again we need to identify what spacetime notions and covariance properties does the deformed algebra for the holonomy corrections imply. 

To aid in such an identification it would be convenient to have an alternative solution for the above set of constraints and the equations of motion since classically two such solutions are equivalent (being related to each other by general covariance). We therefore obtain another time independent solution which classically would correspond to the Painlev\'e-Gullstrand gauge. For this we make the choice $N=1$ which leads to $\ep=x$ and $N^{x}=-f_{2}$. Substituting for $N$ in \eqref{solving holonomy ham} we get
\be \label{holonomy kp eq Painleve}
\frac{f_{1}^{2}}{x}+2f_{2}\kp'=\frac{GQ^{2}}{x^{3}},
\ee
This equation is similar to \eqref{holonomy kp eq schwarzschild} except that the r.h.s. in the present case is non-zero. We can solve this equation following the previous method to find
\be \label{holonomy kp painleve}
\kp=\pm\frac{2x^{2p}}{(\gamma\lP^{2})^{p}}\sin^{-1}\left(\frac{c_{1}}{x^{4p+1}}-\frac{GQ^{2}(\gamma\lP^{2})^{2p}}{4x^{4p+2}}\right)^{1/2}.
\ee
Appealing to the classical solution whereby $\kp=-(2GM/x-GQ^{2}/x^{2})^{1/2}$, we choose the minus sign in the above expression and fix $c_{1}=GM(\gamma\lP^{2})^{2p}/2$. With this choice $N^{x}$ is given by
\be \label{holonomy nx painleve}
N^{x}=\left(\frac{2GM}{x}-\frac{GQ^{2}}{x^{2}}\right)^{1/2}\left[1-\frac{GM(\gamma\lP^{2})^{2p}}{2x^{4p+1}}+\frac{GQ^{2}(\gamma\lP^{2})^{2p}}{4x^{4p+2}}\right]^{1/2}
\ee
which gives the correct classical limit $N^{x}=(2GM/x-GQ^{2}/x^{2})^{1/2}$.

With two (phase space) solutions in hand the only task now is to find a suitable mapping from one to the other. However, for the present case this task appears to be much more non-trivial than for the inverse triad corrections. This can be understood by noting that presently we have only \eqref{solving holonomy ham} to solve for $\kp$ and $N$. For any time independent choice for the form of say $N$, we can solve for $\kp$ and it will be a valid time independent solution for the constraints and the equations of motion. This would also imply that for a given classical solution there are an infinite number of holonomy corrected solutions (in the sense that they all have the correct classical limit). 

For instance, above we found a solution for the choice $N=1$ and noted that in the classical limit this would correspond to the Painlev\'e-Gullstrand solution. But we can as well choose $N=\cos((\gamma\lP^{2})^{p}\kp/x^{2p})$ and to be specific we choose $p=1/2$. In the classical limit, $\sqrt{\gamma\lP^{2}}/x\rightarrow0$, and this choice will also go over to the Painlev\'e-Gullstrand gauge $N=1$ (unless the solution of \eqref{solving holonomy ham} for $\kp$ has some singular behavior in that limit). For the choice $N=\cos(\sqrt{\gamma\lP^{2}}\kp/x)$ however no such singular behavior occurs and the classical limit infact gives $N=1$. 

Thus we have the situation where corresponding to a given classical solution, there are (presumably) an infinite number of holonomy corrected solutions. It is therefore not clear that how or which two quantum solutions corresponding to two different categories of classical solutions (equivalently two different classical gauge choices) should be related to each other. To be specific, since both the choices $N=1$ and $N=\cos(\sqrt{\gamma\lP^{2}}\kp/x)$ have the same classical behavior (and there would be many more such choices) it is not at all easy to decide as to which of these solutions should be related to the solution in equations \eqref{holonomy n schwarzschild}-\eqref{holonomy nx schwarzschild} (with $p=1/2$ for the case under discussion) which classically would correspond to the Reissner-Nordstr\"om solution in Schwarzschild-like coordinates. 

By the same token there would be many quantum corrected solutions which would correspond to the classical Reissner-Nordstr\"om solution in Schwarzschild-like coordinates (where $N^{x}=0$). Let us note that this degeneracy is not related to a possible non-applicability of Birkhoff's theorem for quantum corrected equations. The degeneracy is related to gauge freedom and is present in the classical theory as well. For the classical case (as also for version II of inverse triad corrections where also the constraint algebra is deformed), static gauge $N^{x}=0$ is a valid and unique choice (even a small departure from $N^{x}=0$ implies that the gauge is not static). 

For holonomy corrections $N^{x}=0$ leads to classical solutions (which we therefore ignore) and to obtain non-trivial solutions we can only demand that the gauge choice should go over to $N^{x}=0$ in the classical limit. Since there are many such choices it leads to degeneracy (which would be present even for version II of inverse triad corrections if we impose the weaker condition that $N^{x}=0$ only in the classical limit). 

It is likely that with more quantum corrections included simultaneously, say the inverse triad corrections, this degeneracy will get broken and one will recover unique correspondence between the quantum and the classical solutions. Compared to their homogeneous counterparts where inverse triad corrections are found to be sub-dominant, this further highlights the significance of inverse triad corrections in the context of inhomogeneous systems though in a slightly different context. 

In the absence of suitable notion of a metric, it is unfortunately not feasible to explore the effecs of holonomy corrections on the properties of classical Reissner-Nordstr\"om black hole. We however note that as pointed out in \cite{grain signature, martinpaily1}, with holonomy corrections there is a possibility of signature change. Indeed, with $f_{2}$ as given in \eqref{f2}, the deformation factor $\partial f_{2}/\partial K_{\phi}$ in \eqref{hh bracket phase space holonomy} is
\be
\frac{\partial f_{2}}{\partial K_{\phi}}=\cos\left(\left(\frac{\gamma\lP^{2}}{\ex}\right)^{p}\kp\right).
\ee
For the solution in the Schwarzschild-like coordinates, for instance, using \eqref{holonomy kp schwarzschild}, this leads to
\be
\frac{\partial f_{2}}{\partial K_{\phi}}|_{\text{sch}}=1-\frac{2(\gamma\lP^{2})^{4p+1/2}}{x^{4p+1}},
\ee
which is negative for $x<2^{1/4p+1}(\gamma\lP^{2})^{1/2}$ and would lead to Euclidean signature in the deep quantum regime. Similar conclusion will hold for other gauge choices also (signature change being a gauge independent concept), though in the absence of a suitable notion of covariance two different gauges cannot be directly compared.

%In each case the deformation factor can be negative implying that the signature changes from Lorentzian to Euclidean. However, it is also clear that this condition is not gauge invariant -- the condition for signature change being independent of the mass $M$ and charge $Q$ in the Schwarzschild gauge while explicitly depending on these two parameters in the Painlev\'e-Gullstrand gauge. This implies that while for the Schwarzschild gauge signature change always occurs in the deep quantum regime, for the Painlev\'e-Gullstrand gauge this can occur on classical scales also as for instance when $Q=0$, $M=M_{\odot}$, corresponding to the Schwarzschild solution, the signature changes for $x<(2G\gamma\lP^{2}M_{\odot})^{1/3}$ (where we have put back the factors $G$ and $\gamma\lP^{2}/2$). For massive black holes this would happen at even larger radius which is clearly unphysical. This highlights the neccessity for a guage invariant formulation of the signature change condition also. 

%Thus it seems that one needs deeper insights to decode the covariance properties encoded in the deformed constraint algebra to arrive at suitable spacetime notions. In such a situation it would be much easier to understand the meaning of deformed constraint algebra.
  
%{\bf figures for some of these functions eg. signature change and/or for $N^{x}$ in different gauges? Renormalization?} 

%{\bf The $Q\rightarrow0$ limit in all the solutions would give the results for the Schwarzschild black hole. Inclusion of magnetic charge whether it is feasible at this stage?}

\section{Conclusions}
We considered the effect of loop quantum gravity corrections on the properties of Reissner-Nordstr\"om black holes. Effects of both, the inverse triad corrections and point holonomy corrections, were considered. It turned out that for both kinds of corrections an anomaly-free version of constraint algebra is possible. While for inverse triad corrections, version keeping the constraint algebra unmodified can be found, for holonomy corrections considered here this does not seem to be a possibility. As a further contrast between these two kinds of corrections, we found that for inverse triad corrected constraints and equations of motions, static solutions are possible generically (that is, independent of the specific form of correction functions $\bar{\alpha}$ and $\bar{\bar{\alpha}}$) whereas for holonomy corrections static solutions do not exist and only stationary solutions are allowed. 

As has already been shown in \cite{modifiedhorizon}, we once more confirmed (in the presence of Maxwell field) that for the version of inverse triad corrections resulting in a modified constraint algebra, the spacetime covariance property is modified though this is not the case where only the dynamics is modified and not the algebra. It is a non-trivial task to identify what a modified version of spacetime covariance could be so that one could meaningfully talk about black holes and their horizons. For inverse triad corrections, we were lucky and could identify such a notion based on a \emph{quantum} version of mapping from phase space to spacetime. Since in the quantum theory there is no spacetime to begin with, a possibility where the classical mapping emerges only in the classical limit cannot be ruled out. As required, this quantum version of the mapping goes over to the classical mapping when the quantum gravity effects are absent. 

For the case of holonomy corrected constraints, however, we could not find a modified notion of spacetime covariance. A possible source for this difficulty was pointed out by showing that corresponding to a classical solution in a given gauge there correspond (possibly) infinite quantum corrected solutions which have the correct classical limit. It therefore becomes highly non-trivial to relate quantum solutions in two different gauges. 

However, it is possible that such a large degeneracy is present because we ignored other corrections. It is very much possible that including inverse triad corrections simultaneously with holonomy corrections would lift the degeneracy. This further highlights the significance of inverse triad corrections in inhomogeneous situations compared to their cosmological counterparts. Unfortunately, for this reason we could not write a spacetime metric for the phase space solutions found for holonomy corrected constraints. However, we found that as recently suggested in the literature, there exists the possibility of signature change with holonomy corrections. 

For both versions of inverse triad corrections, however, we could write a covariant metric and analyze some of its properties. We found that while for the case of unmodified algebra quantum gravity corrections resulted in mass, charge and wavefunction renormalization, for the case with modified constraint algebra only the charge was renormalized. A possible wavefunction renormalization (as suggested by phase space solution for the lapse function) did not appear in the metric because of the modified mapping used to go from phase space to spacetime. 

In both the cases, it turned out that Newton's constant was not renormalized. Such a conclusion would have been difficult to draw from the quantum version of vacuum Schwarzschild solution since there the mass is multiplied with Newton's constant and a priori (at the effective level) it cannot be decided whether the correction function is mass renormalization or is it the renormalization of Newton's constant. The presence of Maxwell field breaks the conundrum and shows that $G$ is not renormalized. We also found that inverse triad corrections lead to the possibility of three horizons over a finite range of mass in the deep quantum regime and complete disappearance of the inner horizon (one of the distinguishing features of classical Reissner-Nordstr\"om blackholes) beyond a certain value of the mass parameter.

It is worth noting that with quantum corrections incorporated, it is not guaranteed that classical no hair theorems will hold. One possible future direction of work could be to study quasi-normal modes for the case of inverse triad corrected solutions and see if the quantum corrected solutions are stable against perturbations \footnote{Author is thankful to one of the referees for the suggestion.}.

\section*{Acknowledgements}
The author is thankful to Martin Bojowald for useful discussions and for suggesting improvements to the draft and to Ghanshyam Date for useful discussions. Discussions with Saumitra SenGupta during COSGRAV 2012 which led to this work are kindly acknowledged. This work is supported under Max Planck-India Partner Group in Gravity and Cosmology.

\end{document}